\documentclass[12pt,preprint]{aastex}

\shorttitle{$0.5<z<1$ Field Galaxies with AO/HST}
\shortauthors{Steinbring et al.}

\begin{document}

\title{Keck Adaptive Optics Imaging of $0.5<z<1$ Field Galaxies from the Hubble Space Telescope Archive\altaffilmark{1}}

\author{E. Steinbring\altaffilmark{2,3}, A.J. Metevier\altaffilmark{4, 5}, Stuart A. Norton\altaffilmark{6}, L.M.
Raschke\altaffilmark{4}, D.C. Koo\altaffilmark{4}, S.M. Faber\altaffilmark{4}, C.N.A Willmer\altaffilmark{4,7}, J.E.
Larkin\altaffilmark{8}, T.M. Glassman\altaffilmark{8}}

\altaffiltext{1}{All authors are affiliated with the Center for Adaptive Optics.}
\altaffiltext{2}{Center for Adaptive Optics, University of California,
Santa Cruz, CA 95064}
\altaffiltext{3}{Current address: Herzberg Institute of Astrophysics, National Research Council Canada, Victoria, BC V9E 2E7,
Canada}
\altaffiltext{4}{UCO/Lick Observatory, Department of Astronomy and Astrophysics, University of California, Santa Cruz, CA
95064}
\altaffiltext{5}{NSF Astronomy and Astrophysics Postdoctoral Fellow}
\altaffiltext{6}{Department of Computer Engineering, University of California, Santa Cruz, CA 95064}
\altaffiltext{7}{On leave from Observat\'orio Nacional, Brazil}
\altaffiltext{8}{Department of Physics and Astronomy, University of California, Los Angeles, CA 90095}

\begin{abstract}
We have employed natural guide star adaptive optics (AO) on the Keck II telescope to obtain near-infrared ($H$ and $K'$)
images of three field galaxies, each of redshift greater than 0.5. These are among the highest-redshift non-active disk
galaxies to be imaged with AO. Each of the galaxies was chosen because it had been observed previously with the Hubble Space
Telescope (HST) Wide Field Planetary Camera 2 (WFPC2) by others. Our AO images in the near infrared (NIR) closely match both
the depth and high spatial resolution of those optical data. Combining the archival $V$ and $I$ data with our new $H$ and
$K'$ images potentially produces a long wavelength baseline at uniform resolution. The NIR data probe emission well longward
of the 4000~\AA-break at these redshifts, and provide stellar fluxes less contaminated by dust. We fit two-dimensional galaxy
bulge-plus-disk models simultaneously in all bands, and compare stellar-population-synthesis modeling to the photometry of
these separated components. This is an initial foray into combining HST and AO imaging to produce a high spatial-resolution
multi-color dataset for a large sample of faint galaxies. Our pilot program shows that NIR AO images from large ground-based
observatories, augmented by HST images in the optical, can in principle provide a powerful tool for the analysis of faint
field galaxies. However, the AO data $S/N$ will have to be increased, and AO PSFs need to be controlled more carefully than
they were here.
\end{abstract}


\section{Introduction}\label{introduction}

Disentangling the evolutionary histories of bulges, disks, and star-formation regions is a complicated endeavor. As reviewed
by \citet{Wyse1997} bulges are not universally old small ellipticals inhabiting the centers of young disks. De Jong (1996)
and \citet{Peletier1996} have shown that, at low redshift, there is a strong relationship between bulge color and disk color. Their
near-infrared (NIR) $K$-band imaging avoids the confusion caused by extinction and indicates that there is a wide range of
bulge colors, but for a given galaxy the disk and bulge colors are similar. This may imply similar ages and metallicities of
the stellar populations in each galaxy, and perhaps a shared evolutionary history. At high redshift, \citet{Schade1995} find,
in a small sample of $z>0.5$ galaxies with Hubble Space Telescope (HST) $B$ and $I$ imaging, not only ``normal'' blue
galaxies with exponential disks and red galaxies with prominent bulges, but a significant population of bulge-dominated blue
galaxies - the so-called ``blue nucleated galaxies'' (BNG) -  which may be bulges undergoing active star formation. For
galaxies at redshift $z\sim0.5$ in the HDF, \citet{Abraham1999} showed that Wide Field Planetary Camera 2 (WFPC2) $U-B$ and
$V-I$ colors could be used to separate mean stellar population age from extinction. Studying the stellar populations of
$z>0.5$ galaxies would be improved by obtaining broadband color data over an even wider range of wavelengths. Not only would this
help distinguish between age and extinction, but population models (e.g., Bruzual \& Charlot 1993) show that mass-to-light
ratios are much less sensitive to population age beyond 1 micron rest wavelength. Thus, measurements of structural parameters
at NIR wavelengths should give a much truer picture of the stellar mass distribution. For galaxies with $z>0.7$, the $V$
filter desirably samples rest-frame $U$, which is below the 4000~\AA-break. However, NIR colors then become even more
important because to sample rest-frame wavelengths longward of 1 micron requires $K$-band data.

\begin{figure}
\plotone{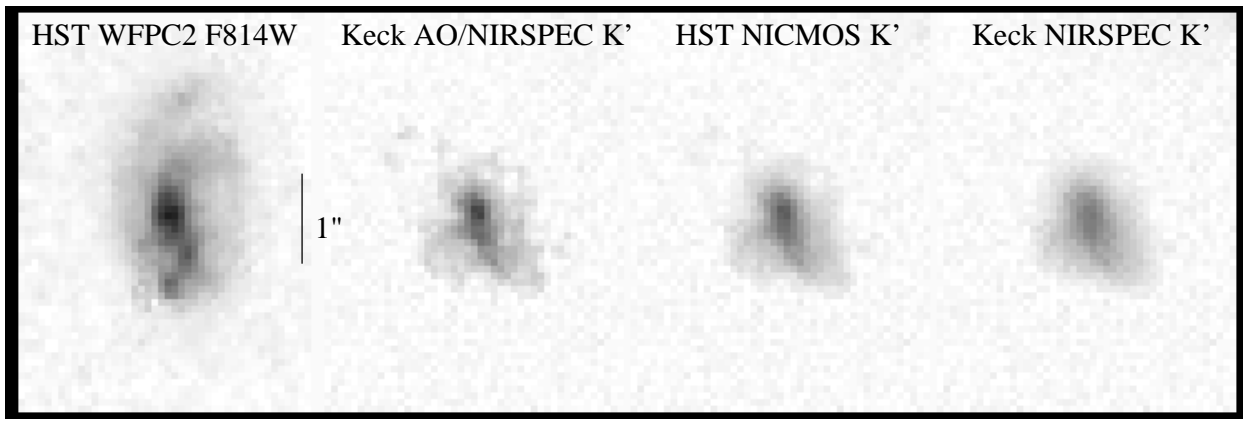}
\caption{Images of JHU 2375 from SA 68. On the left is the WFPC2 F814W image. To the right of this is a Keck AO/NIRSPEC $K'$
image resampled to the WFPC2 image scale. Next to this is the original $K'$ image smoothed to simulate the larger diffraction
pattern for NICMOS. The image furthest to the right simulates the $K'$ image under natural seeing conditions. The simulations
were obtained by smoothing the original ($0.017$ ${\rm arcsec}$ ${\rm pixel}^{-1}$) AO image with Gaussians of FWHM
corresponding to the NICMOS diffraction limit and median seeing for Keck. The pixel sampling of all the $K'$ images has been
degraded to match that of WFPC2.}
\label{figure_jhu2375_comparison}
\end{figure}

To be of any use, such multi-wavelength data must resolve the galaxies spatially, certainly with resolution better than the
typical size of bulges. Low redshift samples (e.g., \citep{Andredakis1995}) suggest that the mean effective radius of bulges might be 
as small as 1.2 kpc. This would correspond to 0\farcs20 at $z=0.5$ ($H_0=70$ km ${\rm s}^{-1}$ Mpc, $\Omega_{\rm
m}=0.3$, $\Omega_{\Lambda}=0.7$; these cosmological parameters will be maintained throughout the text). The data should
ideally cover a large sample of faint galaxies over a significant range in redshift. The only means of obtaining such
high-resolution images in the optical ($V$ and $I$) is with HST. However, the best match to these data in the NIR ($H$ and
$K$) is not provided by the HST NIR camera NICMOS, but rather by ground-based AO. The small aperture of HST means that in the
NIR its resolution is limited to about 0\farcs3. A 10 m class ground-based telescope equipped with AO can in principle achieve
better than
0\farcs1. Figure~\ref{figure_jhu2375_comparison} shows a WFPC2 F814W image of the $z=0.53$ galaxy JHU 2375 and a Keck AO $K'$
image (using the NIRSPEC camera) resampled to the WFPC2 pixel scale (these data will be discussed later in the text). Next to
this is the original $K'$ image smoothed to simulate the larger diffraction pattern for NICMOS (PSF FWHM of 0\farcs3). The image furthest to the
right simulates the $K'$ image under natural seeing conditions (PSF FWHM of 0\farcs5). The resolution and per-pixel $S/N$ in the WFPC2 and AO data
are similar and are both noticeably better than either the non-AO ground-based image or the
simulated NICMOS image. This is a definite advantage for large ground-based
telescopes with AO, but a limitation of Keck and other AO systems (in the absence of laser beacons) is that they can correct only
small regions of sky near bright stars. This precludes for the immediate future a large sample with, say, 1000 faint galaxies
with high-spatial-resolution data in both the optical and NIR.

The Keck AO system requires a $V<12$ natural guide star for optimal correction, which can potentially yield Strehl ratios (the ratio of the guide
star peak to the peak value of a perfect telescope diffraction pattern) better
than 0.4 and provide $0\farcs06$ resolution under good seeing conditions. The AO point-spread function (PSF) has a sharp
diffraction-limited core superimposed on a broad seeing-limited halo. Anisoplanatism dramatically degrades
this on-axis image quality with increasing offset from the quide star, but it is the Strehl ratio and not resolution that suffers most.
The size of the halo changes very little with offset, but as the Strehl ratio worsens the contrast between the halo and the central peak
drops. Typically, the resolution 30{\arcsec} off-axis from the guide star will be about $0\farcs2$, with a PSF Strehl ratio closer to 0.05. Thus for our purposes the combination of anisoplanatism and
guide-star brightness limits the available sky to within about 30{\arcsec} of $V<12$ stars. We cross-correlated the HST
Guide-Star
Catalog and HST Archive to find all WFPC2 pointings that can be reached with Keck that include such stars. The tendency of
WFPC2 observers to avoid bright stars - to prevent saturation and limit scattered light - made these coincidental pointings
rare. We found 38 previous WFPC2 pointings that satisfied this requirement, but only 13 had more than 2000 seconds of
integration, and only two, Selected Area 68 (SA68) and the Groth Survey Strip (GSS), had been targeted by previous galaxy
redshift surveys. Spectroscopic redshifts are essential to establish the true physical size and scale of these objects. As it
happened, both SA68 and GSS had been surveyed by us or our collaborators through the Deep Extragalactic Evolutionary Probe
(DEEP) project. In addition to redshifts, some of our Keck LRIS spectra were of sufficiently good quality to provide mass
estimates from linewidths and rotation curves. Redshifts were in hand for 15 galaxies in SA68 and GSS that were (a) within
30{\arcsec} of the guide-star, and (b) had deep WFPC2 $V$ and $I$ images. Of these, 7 are in SA68, and 8 are in GSS. About
half of these galaxies have redshifts greater than 0.5.

This work involved a pilot program of deep imaging of field galaxies with Keck AO that was designed to provide similar depth
and resolution in the NIR as that provided by HST in the optical. A future sample of hundreds of galaxies will be obtained as
part of the Center for Adaptive Optics Treasury Survey (CATS). The aim of CATS is to use HST's Advanced Camera for Surveys
(ACS) and Keck AO (first with natural guide stars, and then with the laser when it becomes available) to separate galaxy
stellar populations (in the GOODS-North field) into bulges, disks, and star-formation regions and determine relative mean ages.
The addition of the NIR data will provide a means of accurately determining the bulge-to-total flux ratio, $B/T$, of the
galaxy for the mature stellar populations, far less corrupted by dust and recent star formation. For now, we present Keck
natural-guide-star AO
observations of three field galaxies with $0.53<z<0.93$ which had already been imaged with HST WFPC2. In
Section~\ref{data_reductions} the combined HST and Keck dataset and its reduction is outlined. The modeling results are
presented in Section~\ref{analysis} and compared to the available kinematic information in Section~\ref{discussion}. 
Our conclusions follow in Section~\ref{conclusions}.

\section{Data and Reductions}\label{data_reductions}

The HST WFPC2 F606W and F814W data in SA 68 were obtained during the Cycle 6 program 6838 (PI: Kron). The WFPC2 data in GSS
were obtained for program 5090 (PI: Groth) during Cycle 4. Both programs were designed to probe the morphology of faint
galaxies in the optical, and thus deep exposures were made. In SA 68 exposure times were 2400 s in $V$ and 2600 s in $I$; in
GSS they were 2800 s in $V$ and 4400 s in $I$. Structural parameters such as disk scale length $r_{\rm disk}$, inclination $i$,
bulge effective radius $r_{\rm bulge}$, ellipticity $e_{\rm bulge}$, orientation ${\rm PA}$, and bulge fraction $B/T$ have been derived for 
our targets by \citet{Holden2001} (in SA 68) and \citet{Simard2002} (GSS), but this is the first time the WFPC2 images have 
been published. The archival WFPC2 data used in this
work were processed by members of the DEEP collaboration. These images were combined using standard Space Telescope
Science-Database Analysis System tasks and rotated to the proper orientation based on information contained in the image
headers. The signal-to-noise ratio $S/N$ of the HST images is very good. For the brightest structures the per-pixel $S/N$
(assuming noise due solely to Poisson statistics) approaches 100, but is more typically about 20. The fainter structures,
such as disks and possible spiral arms, have $S/N\sim 10$ per pixel. The HST WFPC2 F606W and F814W data were transformed to
standard Johnson $V$ and $I$ according to the prescriptions in \citet{Simard2002}. The point-spread function (PSF) for the
WFPC2 images was determined by measuring a bright but unsaturated star in the field.

\begin{figure}
\plotone{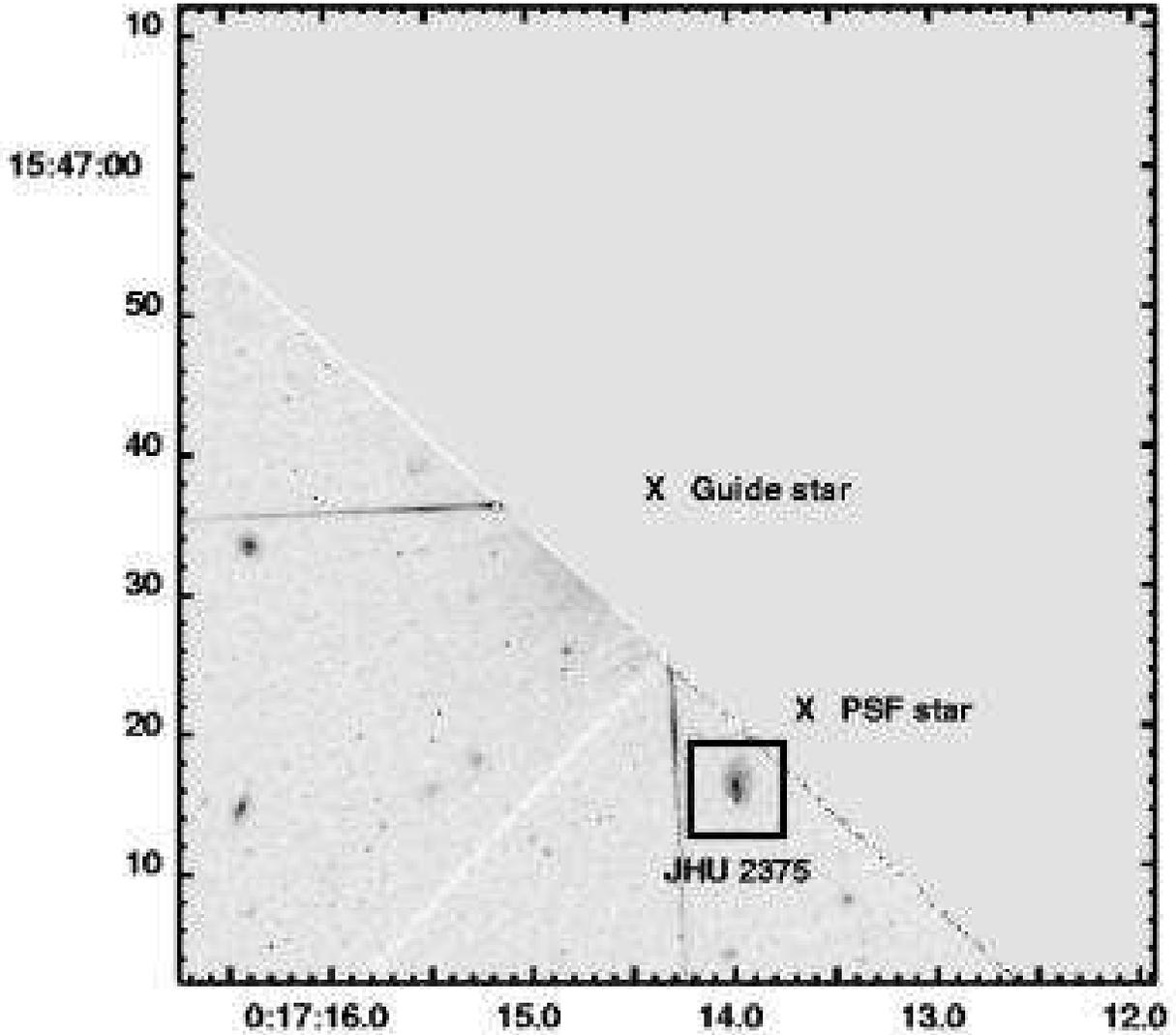}
\caption{The HST WFPC2 $I$-band image of the SA 68 field. North is up and east is left. Right ascension and declination are
given in J2000.0 coordinates. The bright star at the center of this 1{\arcmin} $\times$ 1{\arcmin} field was avoided in the
WFPC2 exposures in order to prevent saturation of the detector. It provided an excellent guide for AO observations. The
position of our $K'$ image is indicated by the dark outline. Another star, fainter than the guide star and closer to the target,
was used to determine the PSF.}
\label{finder_sa68}
\end{figure}

\begin{figure}
\plotone{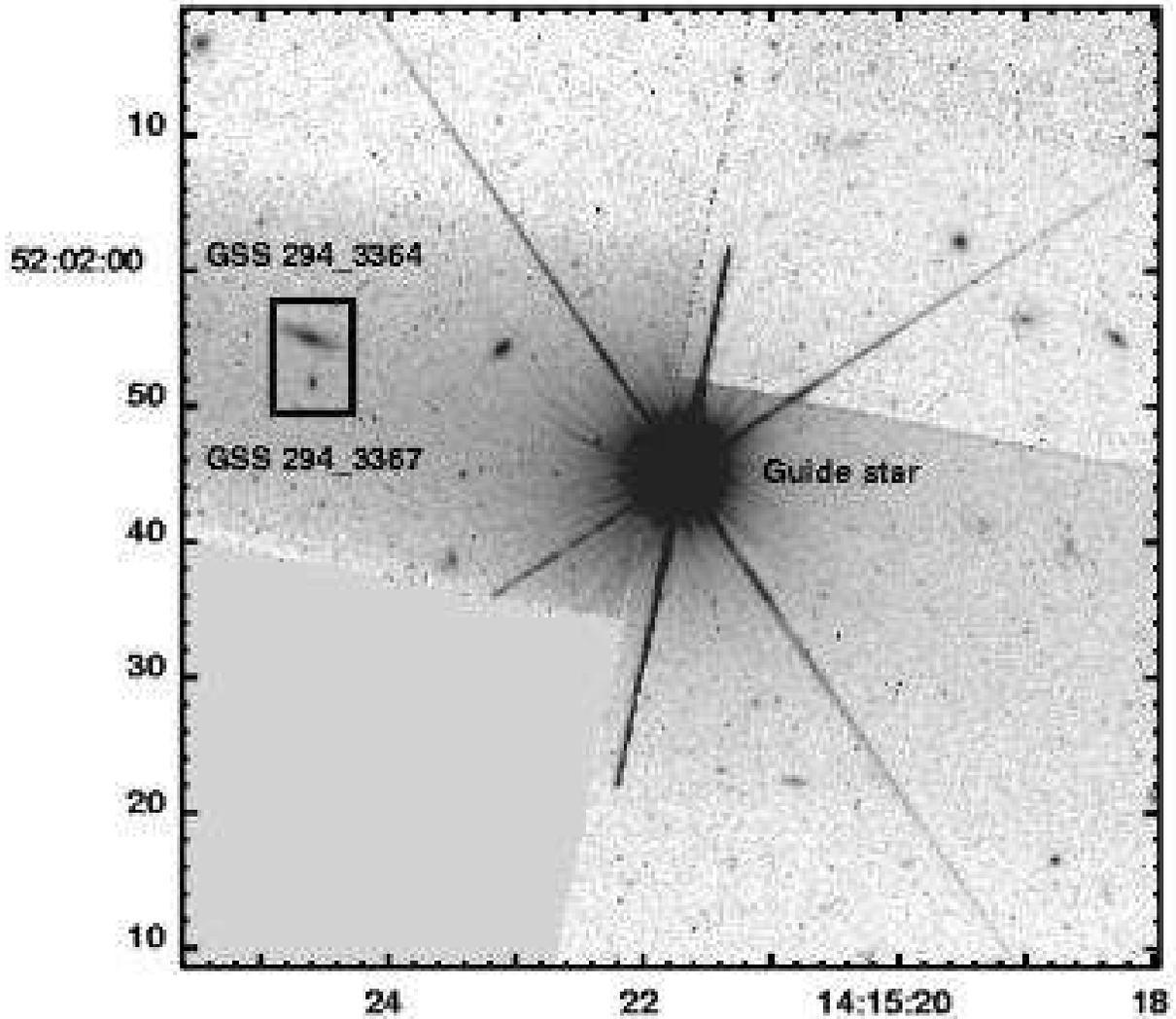}
\caption{The HST WFPC2 $I$-band image of the GSS field. North is up and east is left in this 1{\arcmin} $\times$ 1{\arcmin}
field. Right ascension and declination are given in J2000.0 coordinates. The broad band running diagonally across the image
is due to scattered light from the bright star at the center of the field. The scattered light is fairly uniform - and thus
easily subtracted - at the positions of the targets. We used the bright star as the AO guide. The position of our $H$ and
$K'$ images is indicated by the dark outline.}
\label{finder_gss}
\end{figure}

We obtained Keck II AO observations during observing runs in August 2000 and May 2001. A journal of the observations is given
in Table~\ref{table_journal}. See \citet{Wizinowich2000} for a discussion of the Keck II AO instrument and its performance.
In all observations a $10<V<12$ star was used as a guide. The
uncorrected seeing during our runs was not routinely monitored, but our AO corrected images suggest that it was never much
better than ${\rm FWHM}=$~0\farcs7 in $V$. We made observations through standard $H$ and $K'$ filters. The $K'$ band provides
the largest possible wavelength difference relative to the HST data and should give the best discrimination between young and
old stellar populations. On the other hand, $H$ band should give a better combination of good AO correction and low sky
noise. Because the dedicated NIRC2 NIR camera was not yet available, the NIRSPEC
spectrograph slit-viewing (SCAM) camera was employed as the imager \citep{McLean2000}. This is a $256\times256$ HgCdTe array providing
a $\sim$~4\farcs5 $\times$ 4\farcs5 FOV with 0\farcs0175 pixels. Exposures of either 150 s or 300 s each were made in a
non-repeating rectangular dither pattern with 2{\arcsec} steps. Total exposure times of 2400 s to 3000 s provided images
covering roughly 7{\arcsec} $\times$ 7{\arcsec}. Their locations within the WFPC2 images are indicated in
Figures~\ref{finder_sa68} and~\ref{finder_gss}. In all cases the AO data were resampled to match the 0\farcs0996 pixel
sampling of the WFPC2 data. No smoothing was done prior to resampling. The quality of the AO data is poorer than WFPC2, but
not hugely so. It was greatly improved by resampling to the larger WFPC2 pixel scale. The best per-pixel $S/N$ (at the WFPC2
pixel scale) we obtained for the AO data is 26.5, with more typical values between 3 and 10. Photometric calibration of the
August 2000 data was made by observing standard stars from the United Kingdom Infrared Telesope catalog \citep{Hawarden2001}.
In May 2001 a standard star field in the globular cluster M 5 was used to transform to standard Johnson $H$ and $K'$.

Before we discuss the characterization of our data, a brief introduction to the problem of determining the AO PSF is in order.
A more detailed discussion can be found in \citet{Steinbring2002}. The basic points are as follows: Fainter stars result in poorer correction, 
and for a given guide star brightness the AO PSF is worse with decreasing wavelength $\lambda$, and varies with time and position. 
The best and most stable correction
is obtained at the longest observable $\lambda$. To first order, correction worsens with decreasing Fried parameter $r_0$, which has a
$\lambda^{6/5}$ dependence (see, for example, \citep{Roddier1999}). Thus correction is generally poorer in $H$ than in $K'$. 
Temporal variation in the correction is due to fluctuations in $r_0$, so one must continuously monitor the
variation in PSF with seeing. The spatial variation is due to anisoplanatism, which is inversely proportional to $r_0$. Because
of the $\lambda$ dependence of $r_0$, we would also expect the 
isoplanatic angle to be smaller with shorter wavelength; only $(1.6/2.2)^{6/5}\approx0.68$ as big
in $H$ as in $K'$. Thus for a given $r_0$ the AO PSF is always worse at shorter wavelengths and always worse farther from the guide star.
One further complication is that the AO PSF is also anisotropic, which means that for any
significant offset it is elongated towards the guide-star. Non-radial variation in the PSF as a function of azimuth about the guide star is
expected to be small. Thus it is probably sufficient to simply measure the radial dependence of the PSF and rotate it to account
for any difference in guide-target position angle.

Our observations were all at offsets of approximately 30{\arcsec} from the guide star, where anisoplanatism
significantly degraded the PSF. Thus on-axis calibration observations give an optimistic estimate of the target PSF.
Off-axis PSF measurements can be made at offsets of 30{\arcsec} or more, or made with a shorter wavelength filter than
that used during the science observations - or both. These measurements therefore give a pessimistic PSF estimate.
Together, the on-axis and off-axis PSF measurements provide a
range of possible PSFs, an optimistic ``best-case'' PSF and a pessimistic ``worst-case'' PSF, with the true (unknown) PSF
somewhere in between. 

We obtained on-axis PSF estimates by interleaving target observations with short exposures of the guide star. We began each set
of observations with a measurement of the guide-star PSF, returned to it at 10 to 30 minute intervals, and finished by observing
it a final time. These were combined to determine the time-averaged on-axis PSF.

Off-axis PSF estimates were obtained by observing stars at offsets similar to or slightly larger than those of the targets. For
JHU 2375 (in SA 68), a nearby star was well situated for this task. This star was within 10{\arcsec} of the target and at roughly
the same offset from the guide star (see Figure~\ref{finder_sa68}). It was observed immediately after
target observations were completed (within just a few minutes), but as the seeing conditions were worsening. It therefore gives
an appropriate worst-case PSF estimate. The angular dependence was accounted for by rotating the image to account for the difference
between PSF reference star and target position angles. 
The targets in GSS, unlike SA 68, had no convenient nearby star for determining the target PSF. Thus, calibration observations of
the off-axis PSF had to be taken in a different field, and for practicality, after all of the scientific observations were
complete. To help minimize uncertainty in the temporal PSF variation, the anisoplanatic dependence
was estimated by obtaining a calibrating mosaic image of a crowded stellar field. The value of a mosaic over a sequence of 
on-axis and off-axis measurements is that coincident stars in the small regions of overlap between each pair of
observations could be used to determine if the seeing had varied during construction of the mosaic. The globular cluster M 5 was used.
The small FOV of SCAM required a mosaic with 6 individual pointings to create a strip covering roughly
7{\arcsec} $\times$ 35{\arcsec}. This mosaic was constructed as quickly as possible, in under 10 minutes, hopefully before
seeing conditions could change dramatically and spoil the calibration. The overlap regions confirm that the temporal variations were small. The
mosaic was
observed using a guide star of brightness similar to the GSS observations ($V=10$), and at a similar airmass. We used an $H$
filter in order to record the worst-case PSF. A star at an offset slightly larger than that of the target observation was used as
an estimate of
the target PSF. Again, the angular dependence was accounted for by rotating the image to the same orientation as the target
observations.

The FWHMs of the AO and HST PSFs were determined by fitting them with a two-dimensional Gaussian using the IRAF task IMEXAM.
Our off-axis AO data have modest Strehl ratios. As with uncorrected ground-based data, a Gaussian is a reasonable model for these PSFs.
Our higher Strehl ratio on-axis AO PSFs have a more centrally concentrated core, but retain the broad halo. Thus a Gaussian slightly
overestimates the FWHM of those PSFs. There is probably a similar bias towards overestimating the FWHM of the HST data as the WFPC2 PSF 
also has a sharp core and broad halo, although in this case the halo is from scattered light in the optics \citep{Holtzman1995}. For either
the AO or HST data it is
better to
overestimate rather than underestimate the PSF width for a conservative estimate of whether a given image 
resolves a galaxy. The measured PSF FWHMs are
presented in
Table~\ref{table_resolution}. For the GSS galaxies the resolution of the AO images was between 0\farcs12 and 0\farcs24, a
reasonable match to the 0\farcs17 to 0\farcs20 resolution afforded by HST WFPC2. For JHU 2375, however, the AO PSF may have
been as poor as ${\rm FWHM}=$~0\farcs31.   

\begin{figure}
\plotone{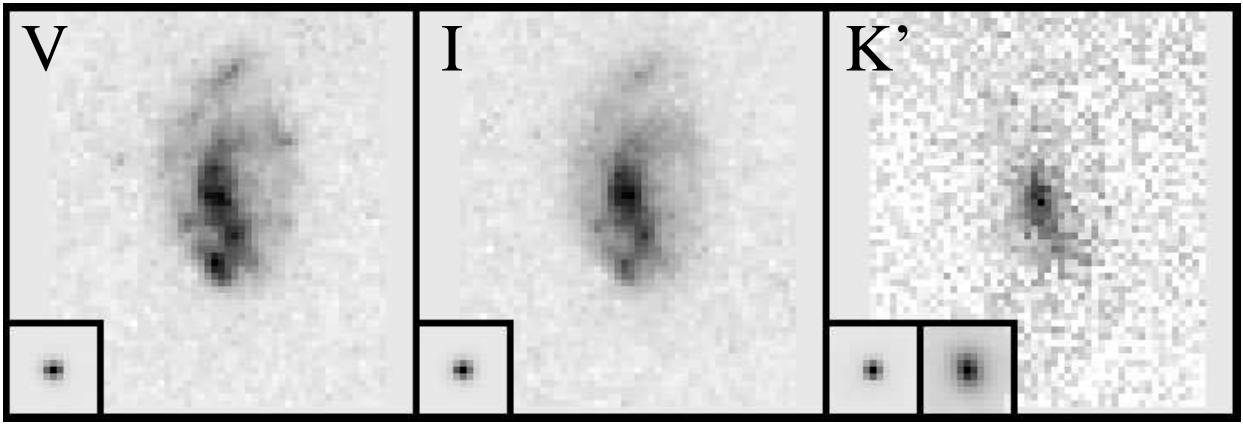}
\caption{Images of the JHU 2375 field ($z=0.53$). North is up and east is left. Right ascension and declination are given in
J2000.0 coordinates. The FOV for each of the panels is 6{\arcsec} $\times$ 6{\arcsec}. The PSFs for the HST data are shown
inset in the lower left corner of those panels. Similarily, the range of possible AO $K'$ PSFs is given - best and worst -  with
the best at far left. See text for details.}
\label{figure_jhu2375}
\end{figure}

\begin{figure}
\plotone{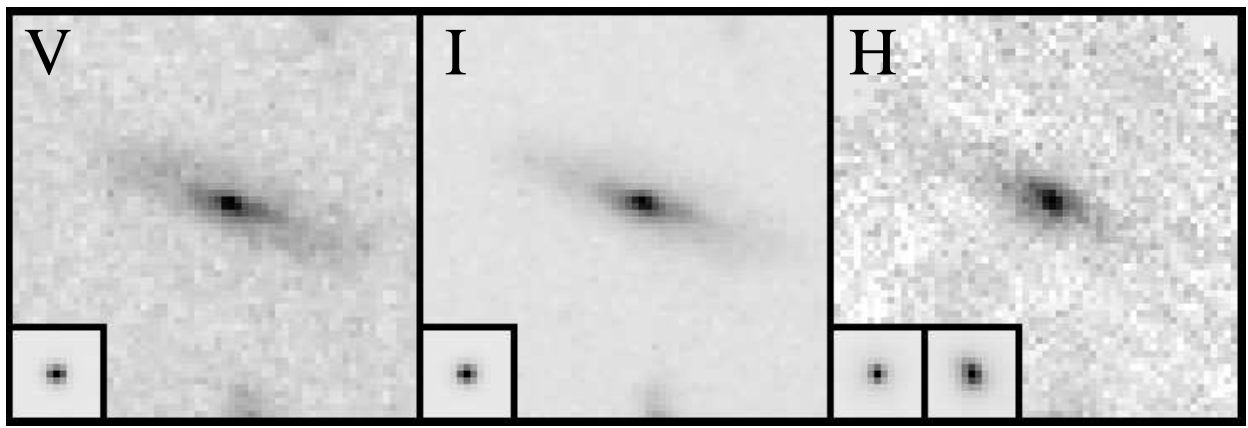}
\caption{Same as Figure~\ref{figure_jhu2375}, but for the GSS 294\_3364 field and replacing $K'$ with $H$. This galaxy has redshift $z=0.65$. The
object
at the bottom of the field is GSS 294\_3367.}
\label{figure_gss_294_3364}
\end{figure}

\begin{figure}
\plotone{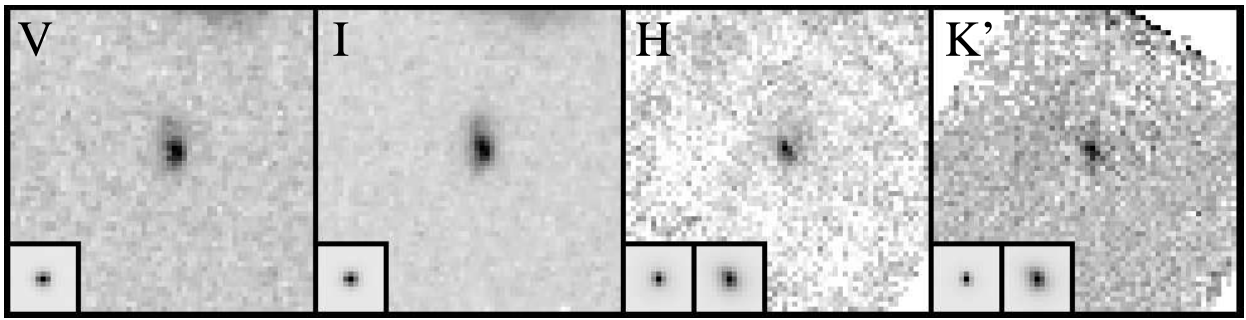}
\caption{Same as Figure~\ref{figure_jhu2375}, but for the GSS 294\_3367 field. This galaxy has redshift $z=0.93$.}
\label{figure_gss_294_3367}
\end{figure}

The images of all three galaxies are shown in Figures \ref{figure_jhu2375}, \ref{figure_gss_294_3364}, and \ref{figure_gss_294_3367}, including the HST
data. Each field is 6{\arcsec} $\times$ 6{\arcsec}, with north up and east left. The best and worst-case PSFs for each filter
are inset in the lower left corner.

\subsection{JHU 2375}\label{jhu2375}

\begin{figure}
\plotone{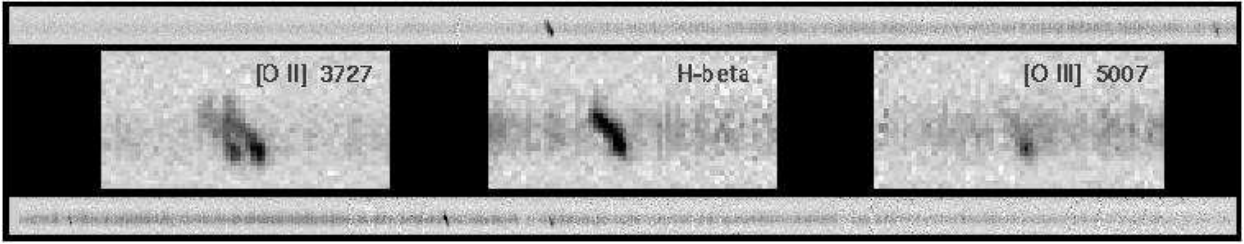}
\caption{Keck LRIS spectrum in the blue (top) and red (bottom) of JHU 2375 ($z=0.53$). In the middle the regions around
prominent emission lines are blown up, clearly showing the rotation curve of the galaxy. Notice the enhanced [O III]
$\lambda5007$ and [O II] $\lambda3727$ doublet emission from the region at the southern tip of the galaxy. The slit P.A. was
20 degrees.}
\label{figure_jhu2375_spectrum}
\end{figure}

The WFPC2 $V$ and $I$ images reveal a great deal of structure in JHU 2375. The redshift of this galaxy is $z=0.53$. The disk
is easily discernible in both images. Two long faint structures to the north may indicate spiral arms. These structures are
also fainter and more diffuse in the $I$ image. In both images an arc of emission extends $\sim1\farcs5$ (about 9.4 kpc in the
rest frame of the galaxy) south from the
bright central core. It terminates in a smaller compact structure, whose bluer $V-I$ (roughly
rest-frame $B-V$) color may indicate a region undergoing active star
formation. The DEEP LRIS spectrum for JHU 2375 seems to confirm this (Phillips, A.C. 2003, private communication). It is
shown in Figure~\ref{figure_jhu2375_spectrum}. The regions around prominent emission lines are enlarged. Each inset is about
4{\arcsec} high and oriented with north up. The H-$\beta$ emission is fairly uniform within the galaxy, but the strong [O
III] $\lambda5007$ line and the [O II] $\lambda3727$ doublet are noticeably enhanced at the site of the southern knot. The rotational
velocity is easily derived from the strong [O II] doublet and is found to be $V_{\rm rot}=144\pm18$ km $\rm s^{-1}$. 

\begin{figure}
\plotone{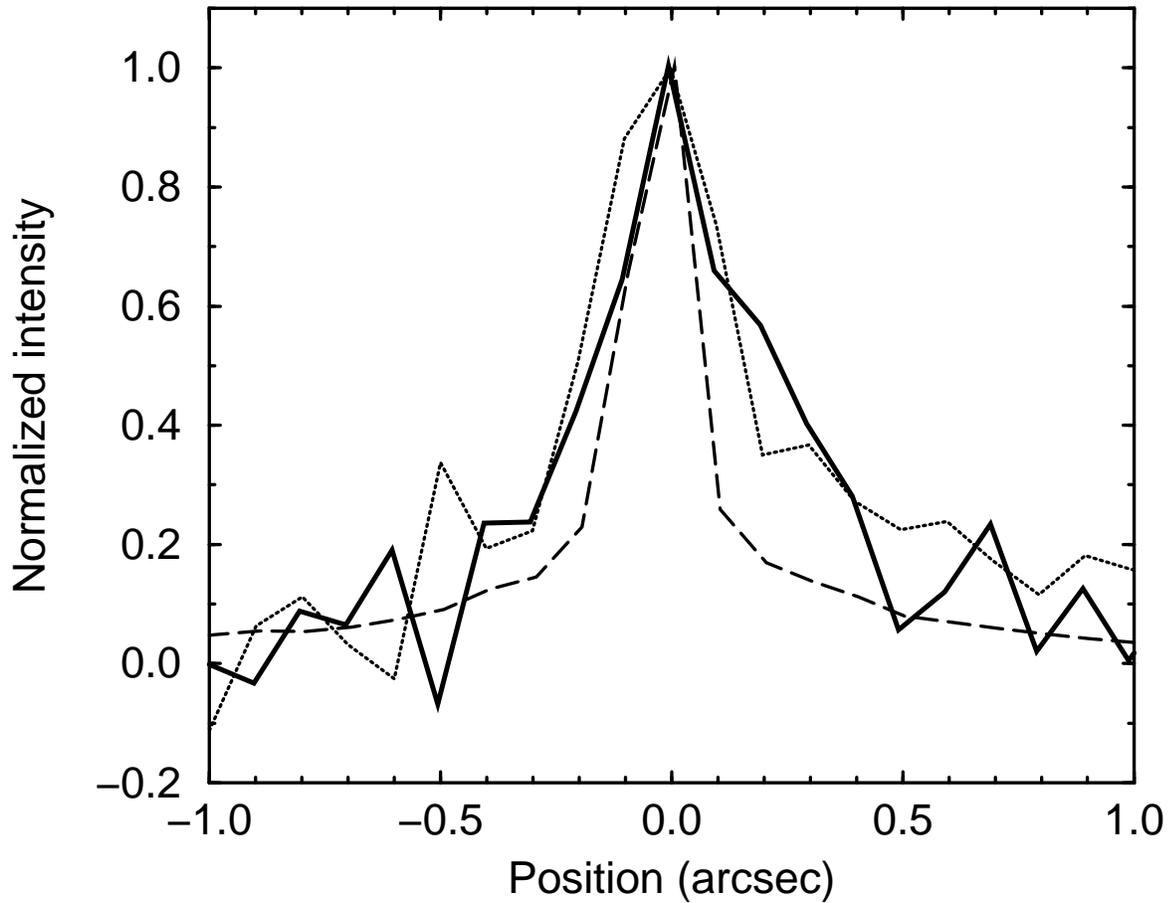}
\caption{The minor-axis $K'$ profile of JHU 2375. The solid line is a cross section running east to west across the center of
the galaxy. Overplotted are the best-case PSF (dashed line) and worst-case PSF (dotted) line. Both have been normalized to the
peak intensity of the galaxy. The width of the best-case PSF is very close to, but slightly smaller than that of the galaxy
profile. The worst-case PSF is an overestimate, and significantly broader than the galaxy peak.}
\label{figure_jhu2375_slice_x}
\end{figure}

Our AO $K'$ image clearly shows that the central region has a compact core. Figure~\ref{figure_jhu2375_slice_x} shows a
minor-axis profile across the galaxy peak in these data, spanning 2{\arcsec} east to west. Overplotted are the best and
worst-case PSF estimates as dashed and dotted lines, respectively. The peak is marginally larger than the best-case PSF, but
the worst-case PSF is probably too broad, and may in fact be broader than the galaxy. The core in our $K'$ image is also elongated
towards the south, consistent with its shape in the HST data. More striking is that the possible star formation region at the
southern tip of the galaxy is not visible here. Its absence may be due to the poorer $S/N$ of our $K'$ data, but if not it
suggests that this population is young and probably not a second nucleus (already unlikely due to the uncomplicated rotation
curve) or an infalling satellite, for example. The faint nebulosity of the disk may also be detected in our $K'$ image but it
is only barely discernible. 

\subsection{GSS 294\_3364}\label{gss_294_3364}

\begin{figure}
\plotone{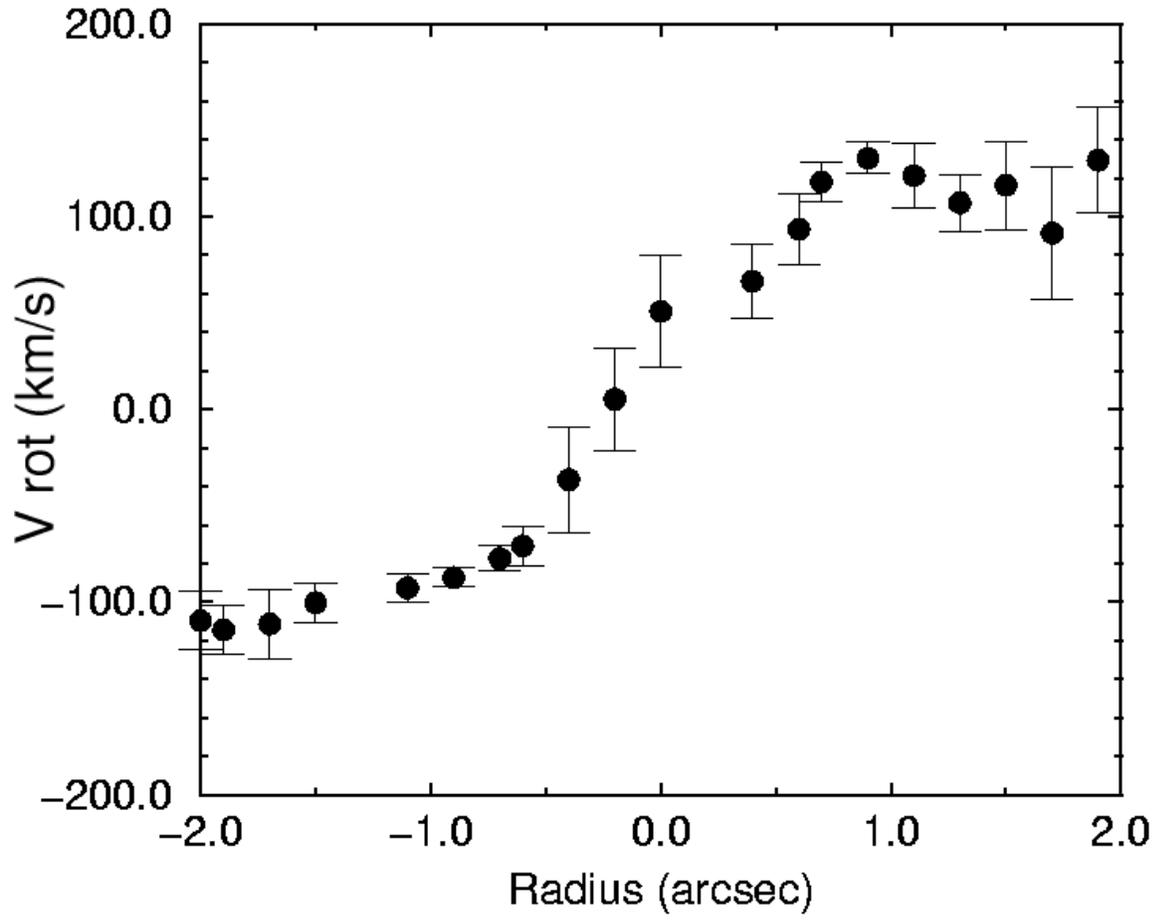}
\caption{Rotation curve of GSS 294\_3364 ($z=0.65$) from the [O II] $\lambda3727$ doublet in the Keck LRIS spectrum.}
\label{figure_gss_rotation}
\end{figure}

The WFPC2 $V$ and $I$ images of GSS 294\_3364 ($z=0.65$) are very similar to each other. They reveal a very smooth nearly
edge-on disk which may be as much as 5{\arcsec} in extent ($\sim 37$ kpc in the rest frame), aligned northeast to southwest. 
This smooth structure is mirrored
in the rotation curve derived from [O II] $\lambda3727$ doublet in the DEEP LRIS spectrum (see Figure~\ref{figure_gss_rotation}).
 After correction for slit position angle (40.5 degrees) and the disk inclination indicated in the WFPC2 images
the derived rotational velocity is $V_{\rm rot}=129^{+23}_{-22}$ km $\rm s^{-1}$ (Vogt, N.P. 2003,
private communication). The HST images also reveal a compact bulge. Our $H$ band AO image resolves this bulge as well, and
also indicates the faint disc running diagonally across the image.

\subsection{GSS 294\_3367}\label{gss_294_3367}

\begin{figure}
\plotone{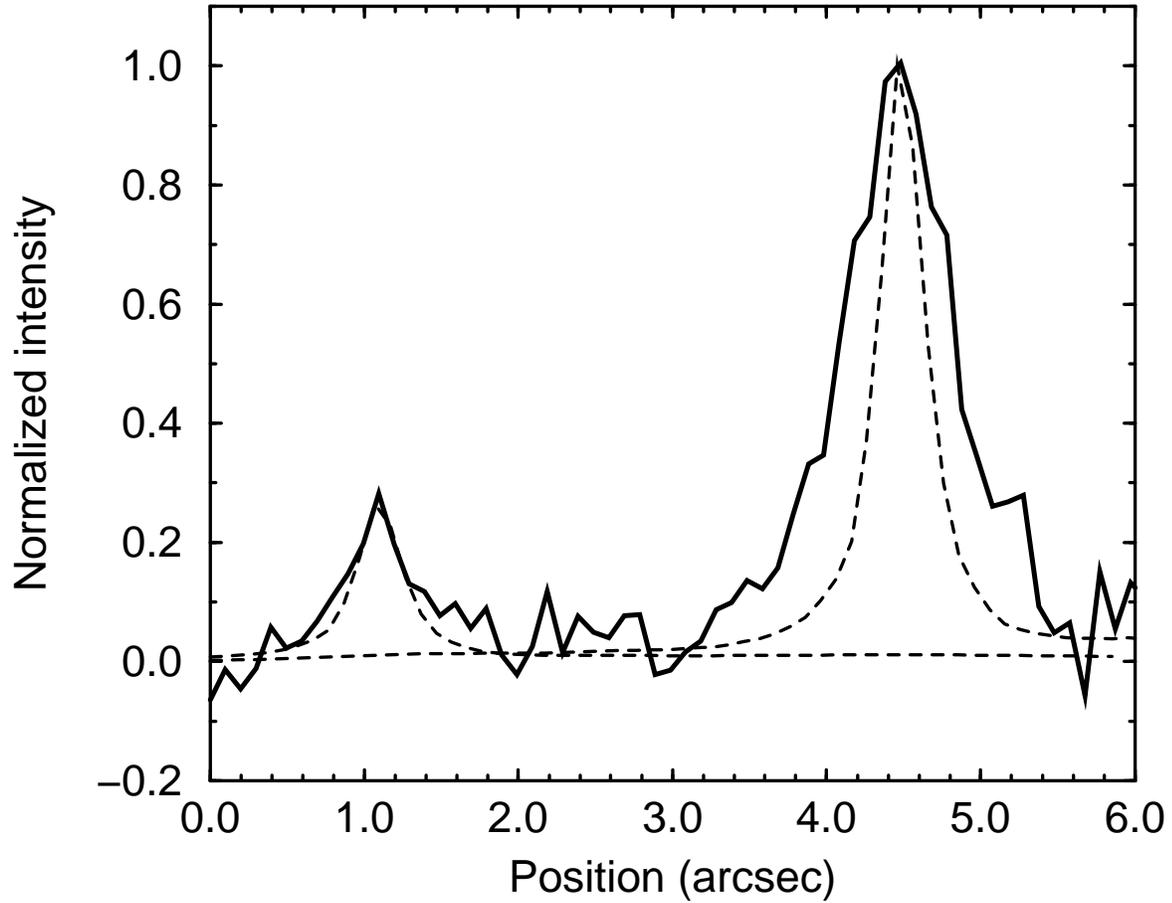}
\caption{The $H$ profiles of GSS 294\_3364 (right) and GSS 294\_3367 (left). The solid line is a cross section running south to north across
the center of the galaxies. Overplotted are the worst-case PSFs. These have have been normalized to the peak intensity of
their respective galaxies. Both galaxies are resolved in our Keck AO imaging, GSS 294\_3367 barely so.}
\label{figure_gss_slice_y}
\end{figure}

The WFPC2 $V$ image of GSS 294\_3367 ($z=0.93$) shows the galaxy to be very small, less than 1{\arcsec} in extent (only 6 kpc in the
rest frame), with a
compact core elongated north-south. The $I$ image is similar, but with some indication of faint nebulosity to the north. Our
$H$ and $K'$ AO images confirm the core to be very compact, spanning only a few pixels. Some of the elongation in our images
may be, in part, an artifact of our noncircular PSF. Both GSS 294\_3367 and GSS 294\_3364 are resolved, however, in the AO
data. Figure~\ref{figure_gss_slice_y} shows a profile of both galaxies. It is a 6{\arcsec} long slice running south to north
across the peaks of each. The worst-case PSFs are overplotted as dashed lines. This plot suggests that even with the
worst-case, broadest PSF, both galaxies would be resolved, if only marginally for GSS 294\_3367. 

\section{Analysis}\label{analysis}

The galaxies were resolved in all of the images. How well resolved is a determining factor in the measurements of $B/T$ and
the photometry, and hence in the relative ages and star-formation histories of bulge and disk. Because we can only set limits on the range of
possible
PSFs, a natural approach is to carry out these analyses in two separate streams: one for best and one for worst PSF. We fit
the data in each band with two-dimensional bulge-plus-disk galaxy models, one model assuming the best case, and the other the
worst case PSF. The output photometry of the bulge and disk components was then compared to the results of simple stellar
poplulation models, guided by the available kinematic information.   

\subsection{Galaxy Structural Parameters and Photometry}\label{morphology}

To determine bulge and disk photometry, we fit each galaxy in our sample with a de Vaucouleurs bulge plus exponential disk
model using GIM2D (Marleau \& Simard 1998, Simard et al. 2002). Although GIM2D can model $r^{1/n}$ profiles for any $n$, we chose
$n=4$ because the $S/N$ of our AO data was too poor to discriminate between other possible $n$ values. This precludes a direct comparison of our
bulge sample with those of local galaxies, which seem to have $n<4$ (e.g.
\cite{Balcells2003}). GIM2D models a galaxy's total flux, $B/T$, bulge and disk
sizes and orientations, and galaxy pixel center from an input image, or simultaneously from multiple images. Bulge-plus-disk
models are convolved with an input PSF and directly compared to data during an optimization with the Metropolis algorithm
\citep{Metropolis1953}. Once the algorithm converges, 99\% confidence intervals are determined via Monte-Carlo sampling of
parameter space.

We first fit GIM2D models simultaneously to $V$ and $I$ band images of the galaxies, as these images typically had higher
signal to noise than those in the NIR filters. This is the same procedure employed by \citet{Holden2001} and \citet{Simard2002} and
reproduces their results for the WFPC2 data. Our version of GIM2D
is not capable of simultaneous fits in three or four bands, so disk sizes $r_{\rm disk}$, bulge effective radii $R_{\rm bulge}$, 
inclinations $i$, and orientations ${\rm PA}$ were then locked to the values determined from the $V$ and $I$
images so that only total flux, $B/T$, sky background values, and galaxy pixel centers were allowed to float in models from
images in other filters. The input PSFs were those described in Section~\ref{data_reductions}. GIM2D fits from WFPC2 images
were run with a single input PSF in each band. The NIR SCAM images were run twice in each band, each with the two different
input PSFs: best and worst. We note that particularly in the cases of the worst PSFs, GIM2D did a poor job of centering the
model, so object centers were determined through an iterative process of fixing centers by hand in several different
positions. The output model with the best reduced-$\chi^{2}$ measurement was taken to have the best-fit center. 

\begin{figure}
\plotone{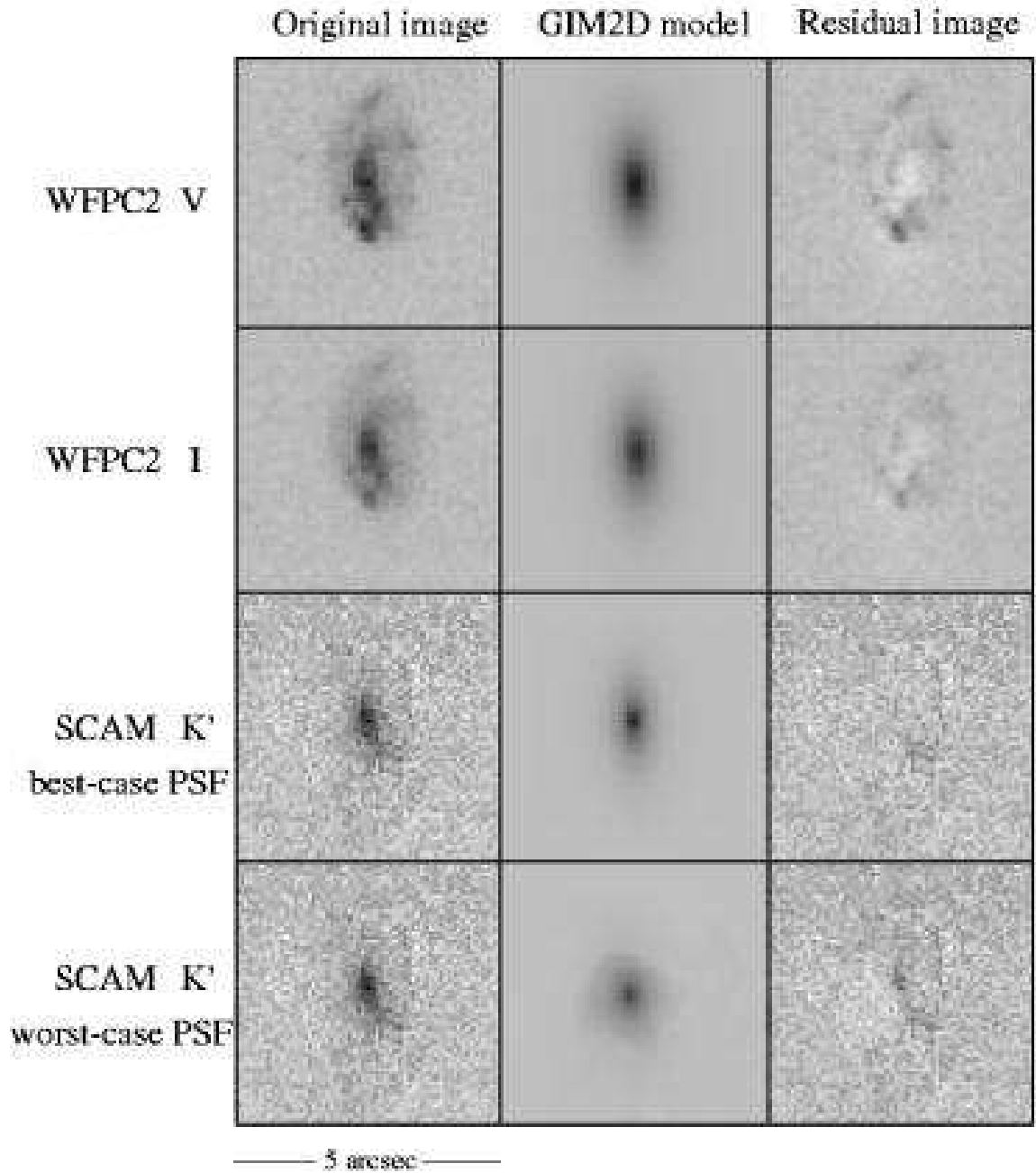}
\caption{Images of the data (left column), models (center column), and residuals (right column) for JHU 2375. Each panel is
5{\arcsec} on a side. The residuals in the WFPC2 images may indicate star formation regions in the disk of this galaxy. For
the $K'$ data, we fit models for both the best and worst case AO PSF, which gives a range of possible output parameters. The
lower $S/N$ of the SCAM data further increases the uncertainty in the $K'$ output parameters.}
\label{figure_jhu2375_fits}
\end{figure}

\begin{figure}
\plotone{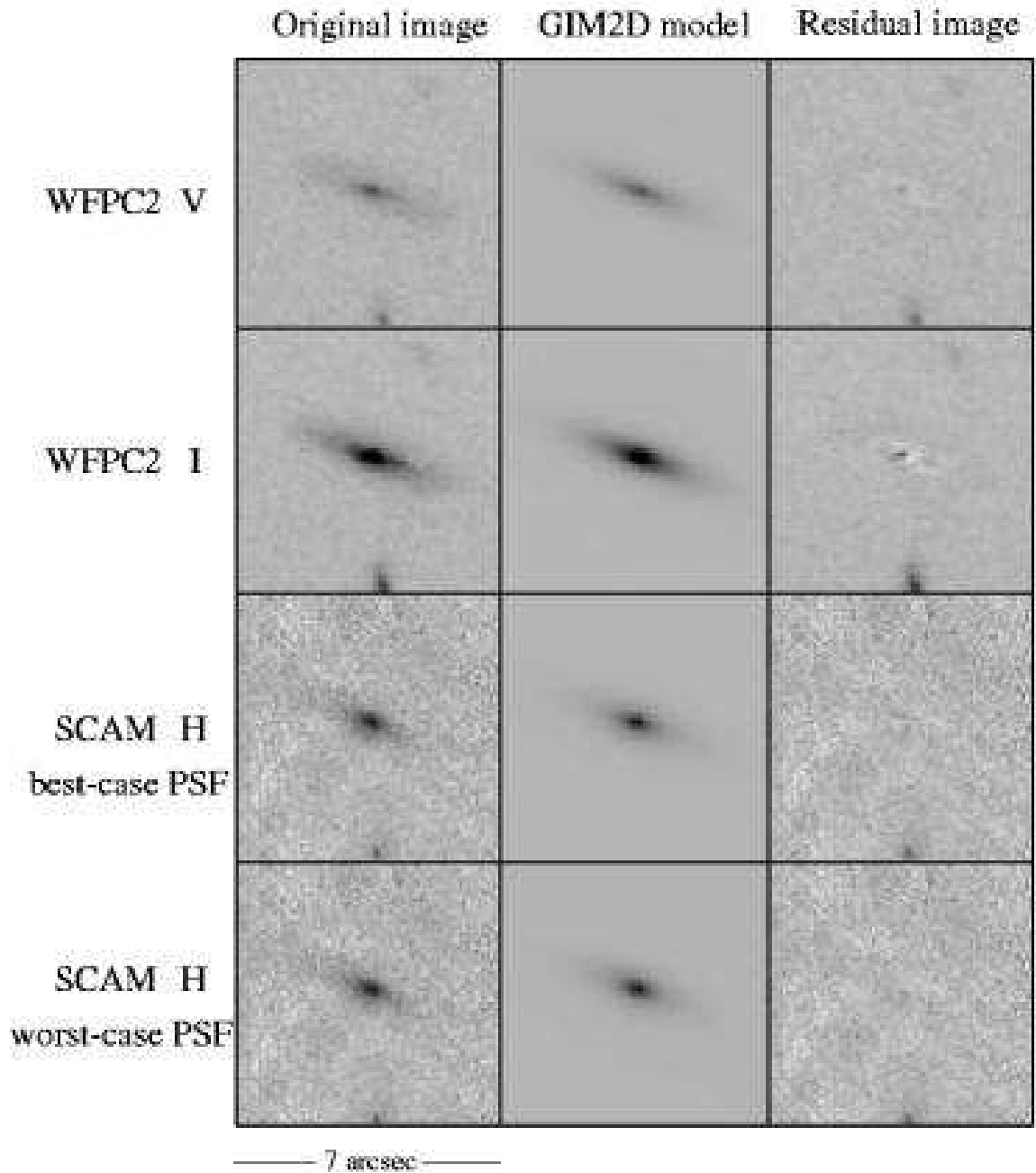}
\caption{Same as Figure~\ref{figure_jhu2375_fits} except for GSS 294\_3364 and for $H$ instead of $K'$.}
\label{figure_gss_294_3364_fits}
\end{figure}

\begin{figure}
\plotone{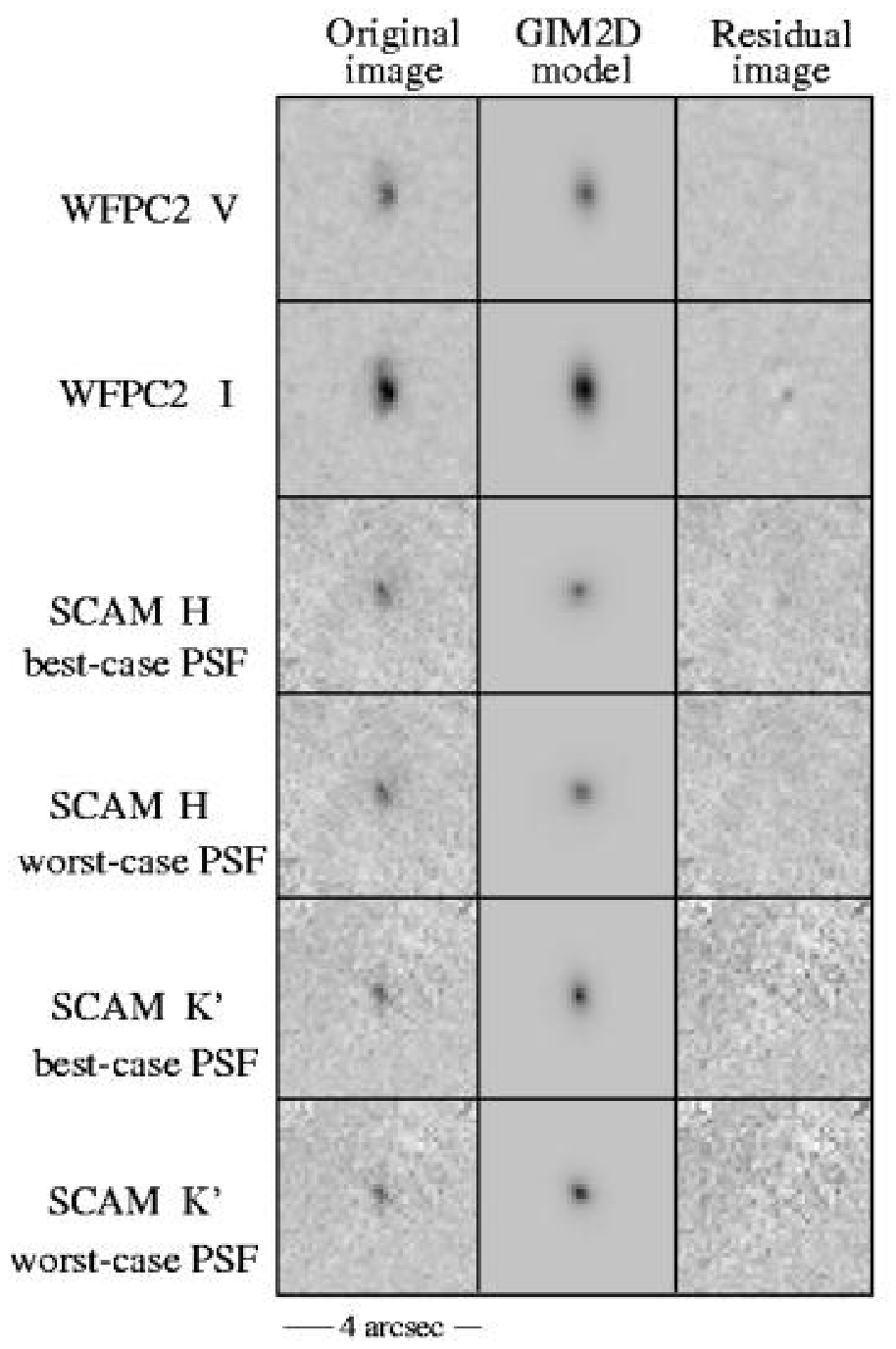}
\caption{Same as Figure~\ref{figure_jhu2375_fits} except for GSS 294\_3367.}
\label{figure_gss_294_3367_fits}
\end{figure}

Figures~\ref{figure_jhu2375_fits} -~\ref{figure_gss_294_3367_fits} show the results. The column of panels on the left is
WFPC2 and SCAM data. Best-fit models for each are displayed in the center column, and the best and worst PSF cases for the AO
data are shown. The residuals are to the right. The WFPC2 PSF is well determined and so the negative residuals from modelling
the $V$ and $I$ images, especially for JHU 2375 (Figure~\ref{figure_jhu2375_fits}) and GSS 294\_3364
(Figure~\ref{figure_gss_294_3364_fits}), may indicate dust in the disks of these galaxies. Notice the prominent positive
residual at the southern tip of JHU 2375. This is the location of the possible star formation region discussed in
Section~\ref{jhu2375}. The amplitude of the NIR residuals allows us to constrain the AO PSF somewhat, at least for JHU 2375. The higher amplitude of
the JHU 2375 worst-case residuals suggests that the best-case PSF is favored here. The residuals for the other two galaxies are in
the noise, which therefore provide no constraint on the PSF.

The resulting structural parameters for all the galaxies are shown in Table~\ref{table_ratios}, quoted with 99\% confidence limits.
The results for the WFPC2 $V$ and $I$ data are those for \citet{Holden2001} (JHU 2375) and \citet{Simard2002} (GSS 294\_3364 and GSS 294\_3367). 
Two of the galaxies (JHU 2375 and GSS 294\_3367) are clearly
disk-dominated in the WFPC2 images, with essentially no bulge being detected above the level of the noise. We fixed our disk
and bulge radii based on the WFPC2 data. It therefore may seem puzzling that we obtain $B/T>0$ in the NIR for JHU 2375 and GSS
294\_3367. But note that even without detecting the bulge the WFPC2 data still provide an upper constraint on bulge size, and
it is the bulge radius which is maintained for the NIR models. For example, the $V$ and $I$ JHU 2375 data permit SCAM $K'$
models with $B/T<0.31$ assuming the best-case $K'$ PSF. If the AO PSF is broader, the model bulge is also broader, and the upper
limit on $B/T$ is relaxed to as much as 0.54 based on SCAM. In Section~\ref{jhu2375} we noted, however, that the true $K'$ could not be much
worse than the on-axis case and so the $K'$ $B/T$ value must be less than about 0.5. The discontinuity in the $B/T$
measurements for JHU 2375, from less than 0.01 to potentially 0.5, is still a concern though. The situation is similar for GSS 294\_3367. Due to the small size
of this galaxy in $H$ and $K'$, if one assumes either the best-case or worst-case $H$ and $K'$ PSFs the bulge and disk models have almost the
same
radius. Because the bulge model falls off more slowly with radius than the disk, this inflates the $B/T$ values to almost 1.
The remaining galaxy, GSS 294\_3364, had significant
$B/T$ values in all bands, and the optical and NIR results were more consistent, especially between $I$ ($0.25<B/T<0.36$) and
$H$ ($0.42<B/T<0.69$). 

Bulges were not unambiguously detected in two of our three galaxies. 
Even so, we attempted to detect any varying position angles of the galaxies' isophotes, especially between the HST
and AO data.  Large deviations between the
bulge and disk PA for a galaxy might be an indication of a bar. This suggests no evidence of bars in the HST data, since for each galaxy the disk
and bulge PAs are consistent.  One might hope that the indications of a bar would be stronger at longer
wavelengths, where such signatures are more prominent. To test this we repeated the NIR bulge-plus-disk decomposition, but allowed both the $B/T$ and PA to
vary. The differences
in returned NIR PA between the initial run and when PA was allowed to float were large ($\sim90$ degrees). However, the uncertainties were equally
large due to the noisy NIR images. Therefore we see can see no evidence of bars from our AO data.

The photometry of the separated bulge and disk components was derived from the total fluxes and the $B/T$ for each galaxy,
assuming best and worst PSFs and the zeropoints discussed in Section~\ref{data_reductions}. Bulge magnitudes therefore correspond to limits that
have been propagated from the total magnitude measurements and their uncertainties. Total, bulge, and disk colors are presented in
Table~\ref{table_photometry}; errors again represent 99\% confidence limits. Our NIR total magnitude of
$K'=16.97^{+0.02}_{-0.01}$ (in Vega magnitudes, which we will maintain throughout) for JHU 2375 is roughly consistent with
the value of $K=17.35\pm0.05$ measured with Keck NIRSPEC without AO by Cardiel (2000, private communication). We do not quote
worst-case PSF photometry for JHU 2375 because this PSF seemed to be broader than the galaxy core (see Section~\ref{jhu2375}).
The label ``knot'' refers to the region of possibly very young stars at the southern tip of the galaxy. Notice that for GSS 294\_3364 and
GSS 294\_3367, assuming a worst-case PSF permits a higher total brightness for the galaxy. This is expected because the
broader PSF wings will make the galaxy model larger and thus it will include more light. If the PSF estimate is too broad
this may provide an overestimate, however, because the broader wings will tend to underestimate sky flux.

Noise in the images and PSF uncertainty both affect the accuracy of the output bulge and disk photometry. To isolate the
effect due only to noise, we ran GIM2D on model galaxy images with artificial noise added to match the noise characteristics
of our WFPC2 and SCAM data. We used the best-fit models as input model images and generated ten noise realizations per
filter. Input PSFs were chosen to match those used to generate the model images. That is, we assumed perfect knowledge of the
input PSF and estimated the error due to noise. We find that output GIM2D-modelled bulge and disk properties match the known
properties of the input model images within the errors, indicating that, for a known PSF, photometric error is attributable
to noise in the image. For the real data we do not know the true input PSF, only a range of possible PSFs, so our resulting
color estimates are bounded by the best and worst PSF values plus their photometric uncertainties.

\subsection{Comparison to Stellar Population Models}\label{population}

We compared an updated version of the \citet{Bruzual1993} population synthesis models (Charlot 1998, private communication)
with the data. We selected the models using the Padova isochrones and a Salpeter initial mass function (IMF) \citep{Salpeter1955} with lower and
upper mass cutoffs of 0.1 $M_\odot$ and 125 $M_\odot$, respectively. These models
include stellar populations with five metallicities: $Z=0.0004$, 0.004, 0.008, 0.02 and 0.05.

From the single-stellar-population models, we can construct spectra for arbitrary star formation histories by adding together
the single-stellar-population spectra from various ages, weighted appropriately. To examine the widest possible range of star
formation histories, we constructed two extreme models: a population that passively evolves (that is, there is no active star
formation) after a brief (100 Myr) burst, and one that maintains a constant star formation rate. Two extreme metallicity
cases for these two models were developed, one metal rich ($Z=0.05$), the other metal poor ($Z=0.0004$). The resulting
spectra represent populations of ages from 1 Myr to as much as 14 Gyr. To construct the $V-I$, $I-H$, and $I-K'$ colors for these
spectra, we shifted the spectra to the redshift of the observed galaxy (which will, of course, restrict the age of the oldest 
possible populations) and then convolved the spectra with the appropriate
filter bandpasses.

In addition to metallicity, star formation history, and age, we need to examine the effects of dust extinction. For this
purpose we used the \citet{Calzetti1997} dust attenuation curve. We calculated the extinction in observed $V-I$, $I-H$, and $I-K'$ and
parameterized our models by $E(B-V)$ color excess.

\begin{figure}
\plotone{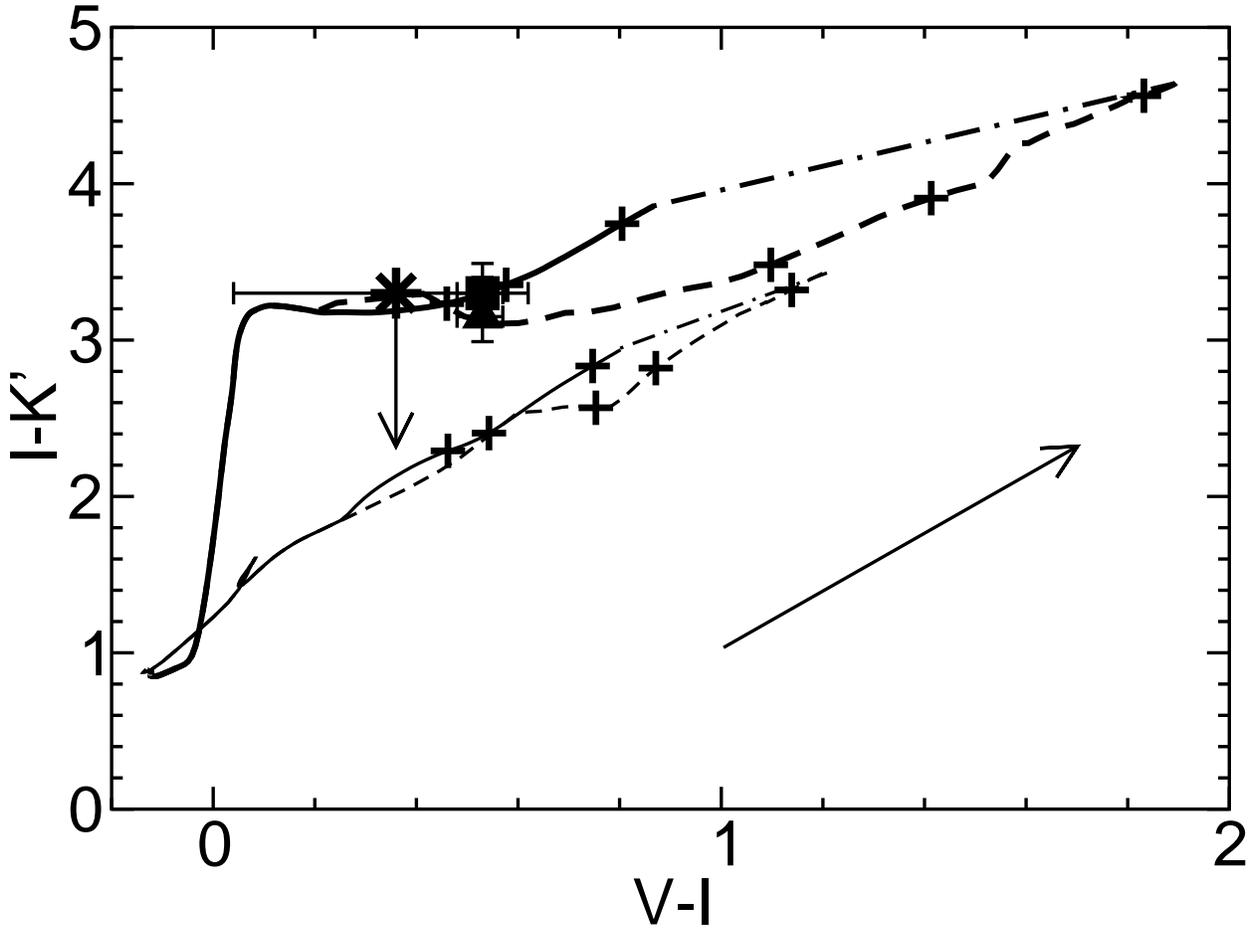}
\caption{Observed $I-K'$ versus $V-I$ color-color plot for JHU 2375.  The total and disk values for the galaxy are indicated
by a filled square and triangle, respectively. The star indicates the knot of possible star-formation. Overplotted
lines are the galaxy isochrones described in the text. Shown are the high-metallicity burst followed by passive-evolution
model (thick dashed line); the same, but with low metallicity (thin dashed line); the high-metallicity
continuous-star-formation model (thick solid line); the same, but with low metallicity (thin solid line). The epoch is
indicated by crosses at 0.5, 1, and 5 Gyr after the onset of star formation. Dot-dashed lines connecting the oldest models
delineate the region encompassed by different star-formation histories; this has a significant effect at high metallicity but
less of an effect at low metallicity. All of the models have been reddened by $E(B-V)=0.5$ in the rest-frame of the galaxy, redshifted
by the $z$ of the galaxy, and observed with terrestrial filters.
An arrow indicates the effect on the models of further increasing dust extinction by $E(B-V)=0.5$.}
\label{plot_jhu2375}
\end{figure}

\begin{figure}
\plotone{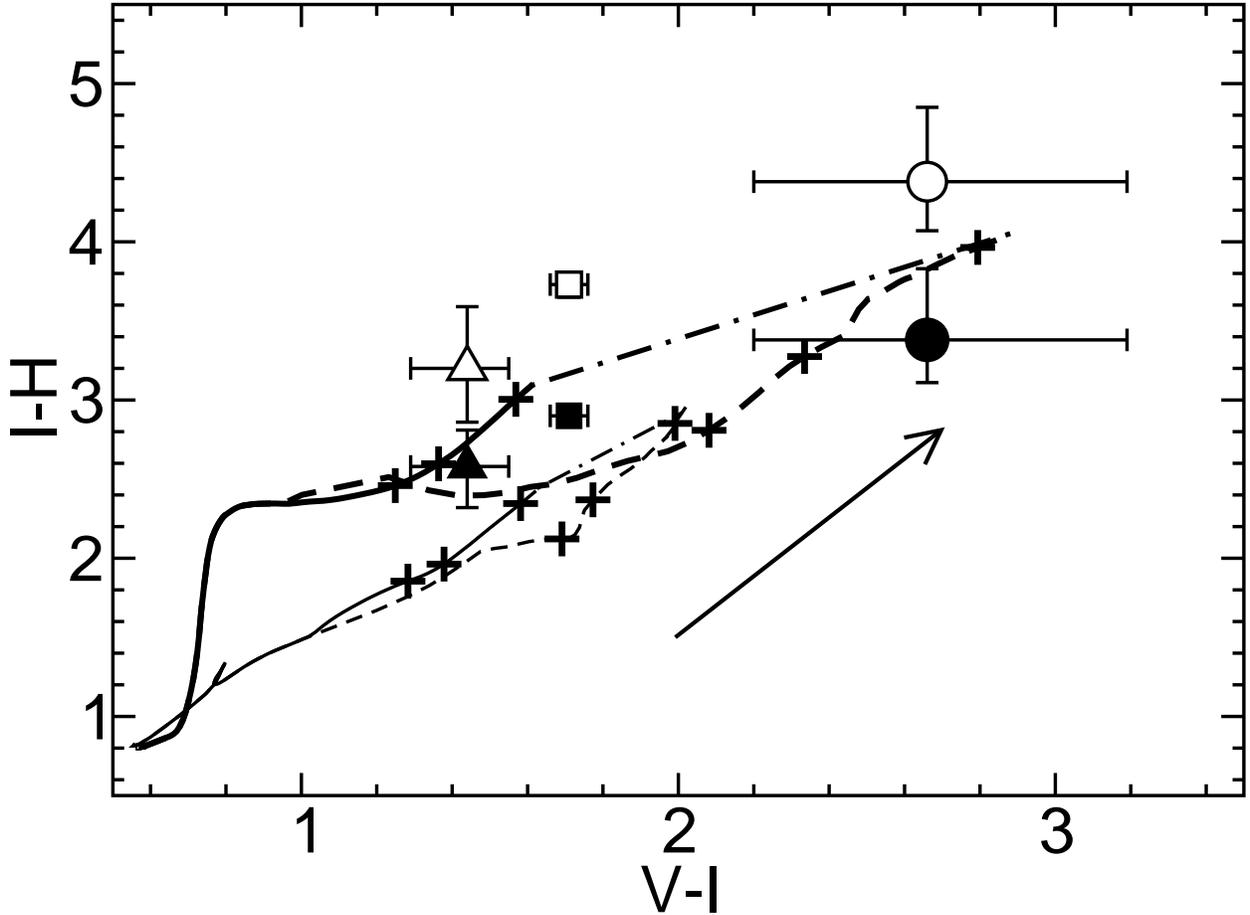}
\caption{$I-H$ versus $V-I$ color-color plot for GSS 294\_3364. The symbols are the same as in Figure~\ref{plot_jhu2375}
except that here a circle represents the bulge.  The results assuming a best-case PSF are shown as filled symbols and
those with a worst-case PSF are indicated by open symbols.}
\label{plot_gss_294_3364}
\end{figure}

\begin{figure}
\plotone{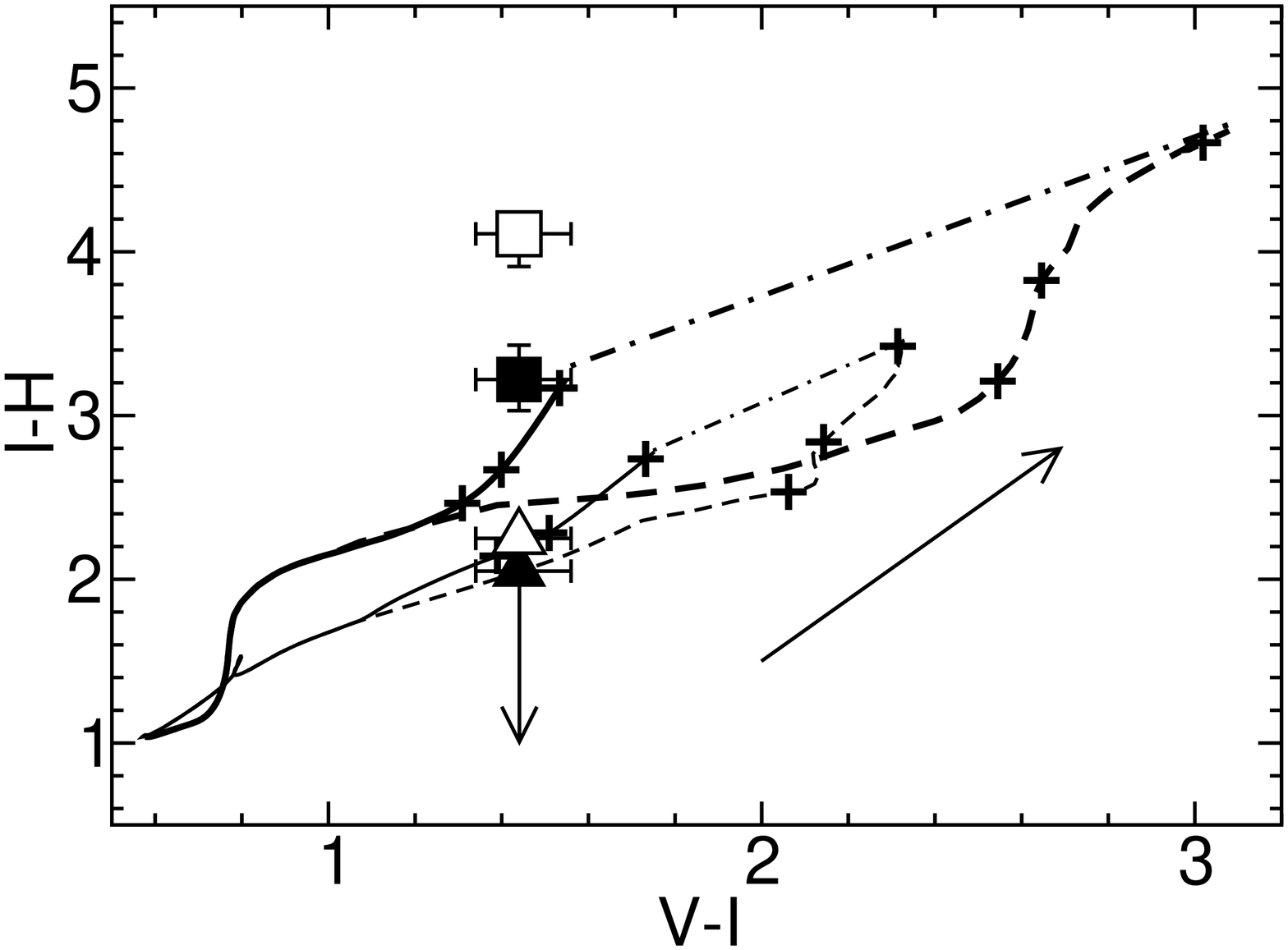}\\
\plotone{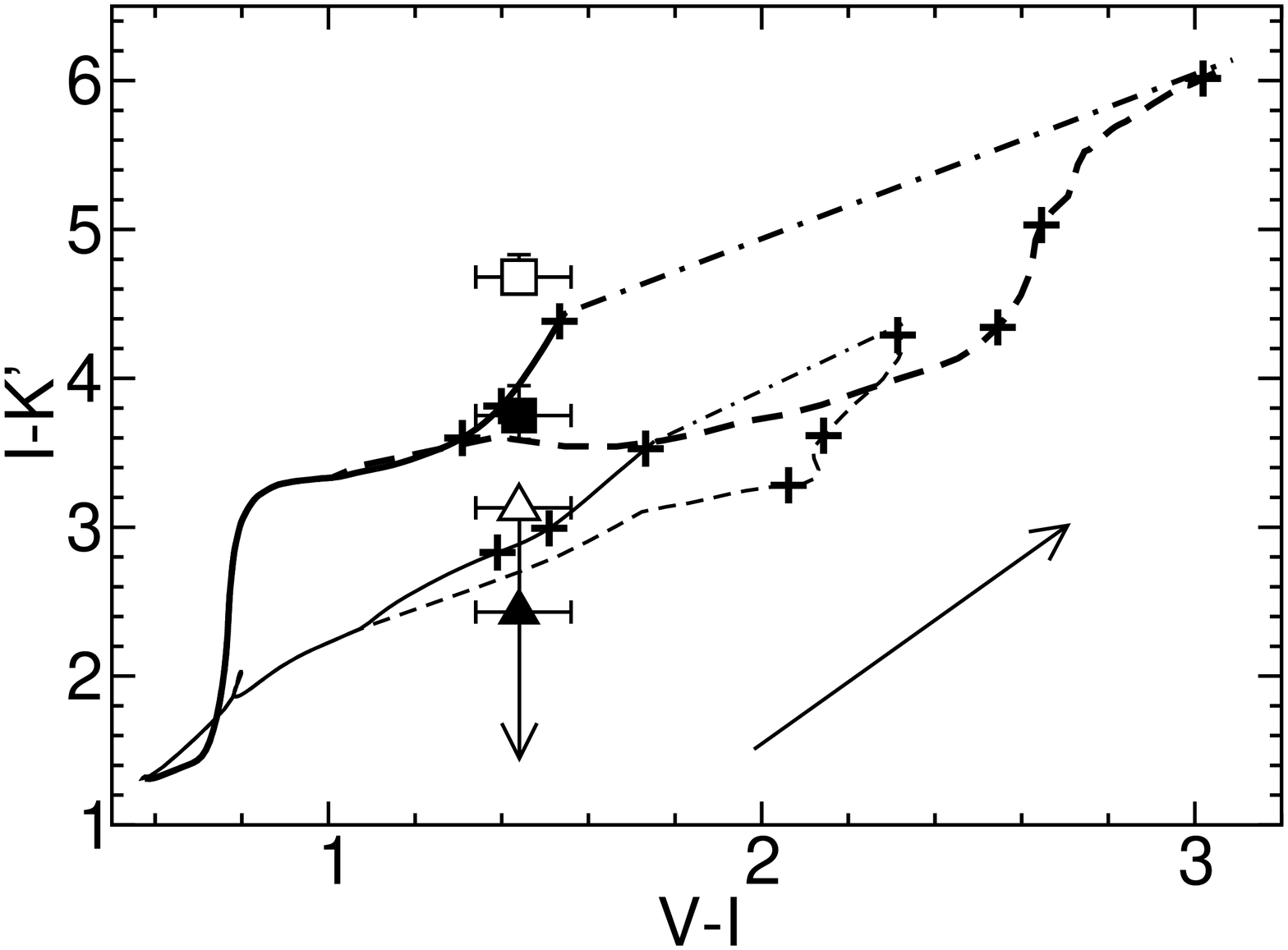}
\caption{$I-H$ versus $V-I$ and $I-K'$ versus $V-I$ color-color plots for GSS 294\_3367. The symbols are the same as in
Figure~\ref{plot_gss_294_3364} except that here no bulge is plotted.}
\label{plot_gss_294_3367}
\end{figure}

The photometry of Section~\ref{morphology} is plotted against these models in Figures~\ref{plot_jhu2375}
-~\ref{plot_gss_294_3367}. The total, disk, and bulge (Figure~\ref{plot_gss_294_3364} only) components are represented by squares, triangles, and circles,
respectively. The data assuming the best-case PSF are indicated by solid symbols, and the data assuming the worst-case PSF by
open symbols (Figures~\ref{plot_gss_294_3364} and~\ref{plot_gss_294_3367} only). We have overplotted the galaxy models. In
each plot we include an arrow indicating the effect of dust extinction with rest-frame $E(B-V)=0.5$. For all the galaxies,
because it appears necessary to explain the observed colors, we apply an extinction correction to the models corresponding to
rest-frame $E(B-V)=0.5$.

Figure~\ref{plot_jhu2375} suggests that the bulk of the stellar population in JHU 2375 is metal rich and young or at least
still forming strongly. The southern knot may be bluer than the rest of the galaxy, which would be consistent with it being a region of active
star formation. Adding more dust to the galaxy will only make it bluer and thus give even younger ages. We have no
constraint on the $V-I$ color of a bulge, and hence very little information about its age or star-formation history. It is equally
compatible with having the same or a very different star formation history relative to the disk.  

The AO photometry of GSS 294\_3364 is better measured than JHU 2375, and therefore it is easier to judge which of the two
main star-formation models is a better fit in Figure~\ref{plot_gss_294_3364}. In fact, it seems that both old and young models are
necessary to describe the separated disk and bulge light if both are to have formed at the same time. The best-case AO PSF seems
to be a better fit to the data. If so, the high-metallicity
continuous-star-formation model matches the total light and disk data well, whereas a passively evolving high-metallicity
model is favored for the bulge. The mean age of the galaxy based on the bulge may be 7 Gyr, but, with added dust, the PSF
uncertainty can accommodate much younger ages - even as young as 1 Gyr. 

The range of allowed integrated $I-K'$ color for GSS 294\_3367 in Figure~\ref{plot_gss_294_3367} makes it difficult to determine
an age for this galaxy. An age as old as 6 Gyr is possible by assuming the worst-case PSF and continuous star formation, but younger ages can be
accommodated by either adding dust to that model or by adopting the best-case PSF and a burst-plus-passive-evolution scenario. Note that the best-case
PSF integrated $I-H$ color for GSS 294\_3367 agrees with only the reddest model predictions, whereas multiple models are in good agreement with the $I-K'$ data.
One plausible solution to this minor discrepancy is to appeal to a combination of the tiny size of this galaxy (only a few pixels across) and the
difficulty of AO PSF
characterization. Correction is better and more stable in $K'$ than in $H$. This suggests that we may slightly overestimate the FWHM of our
``best-case" $H$ PSF. Poorer resolution of the galaxy would tend to predict a redder best-case PSF color, which
is what one might expect by looking at the red worst-case PSF datapoints in this plot. To see if choosing a sharper PSF might give a bluer $I-H$
color we re-ran the models with the best-case $K'$ PSF. This model did not converge; it may be that the better data for GSS 294\_3367 is in $K'$, where the galaxy is
better resolved. The disk $I-K'$ color of GSS 294\_3367 seems to favor low metallicity
and young age.  

The overall result of the stellar population modeling is that the light of all the galaxies' disks could be
described by either a burst$+$passive or continuous star formation history. Typically, most populations are quite metal rich, but
significant variation in the amount of dust can be accommodated, and thus ages are poorly constrained.
For GSS 294\_3364 - the only galaxy with a prominent bulge in all bands - the difference in color between bulge and disk is
plausibly due to different star formation histories. The bulge and disk may have started forming stars at the same time, but
the redder bulge can be explained by its being formed in an initial burst followed by passive evolution, whereas the disk is
still undergoing constant star formation. 

\section{Discussion}\label{discussion}

Our galaxies seem to be typical spirals. 
The integrated WFPC2 $I$ magnitude of JHU 2375 is 20.3, which gives
an absolute rest-frame $B$ magnitude of $-21.1$ using the K-corrections in \citet{Simard2002}. Similarily, $M_B$ of GSS
294\_3364 is approximately $-21.4$ and GSS 294\_3367 has $M_B\approx-20.9$. These rest-frame $B$ magnitudes would correspond
to the high $V_{\rm rot}$ end of the Tully-Fisher relation (TFR) for local galaxies \citep{Pierce1992, Ziegler2002}. 
There is some debate about whether the TFR evolves out to $z\sim1$. \citet{Vogt1997} suggest that there 
is essentially no change in zeropoint or slope over the redshift range $0.1<z<1$ with respect to the local TFR. \citet{Ziegler2002} find a change in slope of TFR 
with redshift, but their TFR for $0.1<z<1$ galaxies crosses the local one at about $M_B=-21$. This suggests that our galaxies 
should have masses about the same as local spirals of the same brightness. For the two galaxies for which we can obtain mass 
estimates from $R_{\rm disk}$ and $V_{\rm rot}$ we find $M_{\rm tot}=6.3\times10^{10}$ $M_\odot$ (JHU 2375) and $M_{\rm tot}=7.2\times10^{10}$ $M_\odot$ 
(GSS 294\_3364). 

The sizes of our two galaxies with rotation curves are somewhat large for their $V_{\rm rot}$, but otherwise within 
the spread of $\log{(R_{\rm disk}/V_{\rm rot})}$ seen in larger samples.
\citet{Mao1998} find that for a given rotation speed, $V_{\rm rot}$, disk scale length, $R_{\rm disk}$, decreases as $\sim(1-z)^{-1}$ out to
$z\sim1$. This suggests
a peak in the high-$z$ distribution of $\log{(R_{\rm disk}/V_{\rm rot})}$ at about $-1.8$. JHU 2375 and
GSS 294\_3364 lie at about $-1.5\pm{0.2}$ and $-1.3\pm{0.2}$ respectively.

Massive galaxies tend to have higher metallicity, and local galaxies with absolute $B$ magnitudes similar to ours
fall in the solar to super-solar range \citep{Zaritsky1994}. The $L-Z$ relation does not seem to evolve much out to $z\sim1$ for $M_B\sim-21$
galaxies \citep{Kobulnicky2003}, which suggests that we should expect ours to have $Z>0.01$. The results of our stellar population
synthesis modeling indicate that our three galaxies are indeed plausibly metal rich, since the colors of the integrated light seem to
agree with our high-$Z$ models. But this last result must depend at least in part on the properties of the bulges, since for the
one target where we unambiguously detect a bulge it may constitute a significant fraction of the mass.

In the local universe, the bulges of spirals seem to follow an extension of the
tight relation between luminosity and masses of low-mass ellipticals. If one includes the bulges of spirals in the $B$-band fundamental plane (FP)
of local galaxies, 
those with integrated $M_B=-21$ should fall at about $2\times10^{11}$ $M_\odot$ and $M/L_{B}\approx6$ \citep{Bender1992}. The $B$-band FP relation
for $0.3<z<1$ galaxies is also tight, with perhaps the most massive galaxies being somewhat brighter in the past for everything else being equal,
including $M/L_{B}$ \citep{Gebhardt2003}. The one galaxy where
we unambiguously detect the bulge is GSS 294\_3364.
If we simply assume for the moment its bulge and disk share the same $M/L$ ($1.8M_\odot/L_\odot$ based on its kinematic mass), our $B/T$
values can then be used to estimate a bulge mass. 
The bulge in rest-frame $B$ is small, certainly with a mass less than 
$7\times10^9$ $M_\odot$. Does this mean that the bulge is still being constructed?  Possibly not because our NIR data are 
compatible with higher masses, perhaps as much as $M_{\rm bulge}=5\times10^{10}$ $M_\odot$. This is in better agreement with the observed masses
of local bulges found by \citet{Bender1992}.

Spirals in the local universe seem to have similar disk and bulge colors, which suggests they probably share the same
age, metallicity, and perhaps even star-formation history \citep{deJong1996, Peletier1996}. The metallicities of our 
galaxies (in integrated light) would seem to be typical of massive local spirals based our stellar population synthesis 
modeling. The color of the GSS 294\_3364 bulge suggests that it shares the same high metallicity as the disk, even if 
it formed in an initial burst rather than still actively undergoing star formation, as the disk seems to be. We caution 
that this is just one galaxy, and the result depends strongly on the allowed $B/T$, and thus knowledge of the delivered AO 
PSF FWHM. This would seem to be an argument for obtaining high-spatial-resolution NIR AO images of a larger sample of 
$z\sim1$ field galaxies, with good control of the AO PSF. The laser beacon on Keck is one path towards obtaining this, 
and a key scientific program of the CfAO, the CATS survey, plans to take advantage of this route.

\section{Conclusions}\label{conclusions}

We have discussed a program of NIR AO observations of field galaxies where archived optical HST data were already available.
The addition of the NIR data provided new information about the star-formation histories of these systems. The high spatial
resolution allowed us to fit two-dimensional galaxy bulge-plus-disk models so that stellar-population-synthesis modeling
could be carried out on the separated components. Two of the galaxies are disk dominated in the optical but possibly not in the NIR:
JHU 2375 and GSS 294\_3367. A significant bulge component is detected in all bands for GSS 294\_3364. The integrated
colors of all three galaxies can be fit by a model with either extended periods of star formation or passive evolution after an initial burst. Some
metal enrichment is needed to fit the data. In all cases, the age of the galaxy is not well determined because of poor
constraints on dust extinction; in turn due mostly to poor constraints on the AO PSF. For GSS 294\_3364, one way to explain
the difference in color between the bulge and disk is that the star formation histories of the bulge and disk are very
different. The bulge evolved passively after an initial burst, and the disk is still undergoing active star formation. 

This is an initial attempt at obtaining multi-color photometry of faint galaxy substructures by providing NIR images at
spatial resolution and depth comparable to HST. We achieved similar resolutions with Keck AO, but our ground-based $S/N$ was
generally poorer. One conclusion of this pilot study is that AO exposures need to be approximately a factor of 5 times
longer than we have obtained in order to properly match the $S/N$ of the HST data. More importantly, PSF uncertainties introduced 
significant errors; these would have to be reduced by at least a factor of two for serious work. These requirements set goals for the next phase of
our
work.  Clearly AO data provides a significant advantage over HST alone for determining the bulge properties of $z>0.5$ spirals. For our one 
galaxy where the bulge
is unambiguously present, the sharp AO PSF nicely resolves this
bright compact structure at a rest wavelength where it is bright. AO does not do as well with low surface-brightness features
such as disks, and thus in this paper we have the HST data to constrain them. We have not investigated whether sufficiently deep seeing-limited data
can provide the necessary information about extended, low surface-brightness emission that the AO data currently do not. 

\acknowledgements

We would like to thank the staff of the Keck Observatory, especially observing assistant Terry Stickel and AO specialist
David Le Mignant. We thank Nicolas Cardiel for his NIR Keck photometry, Drew Phillips for his LRIS spectrum of JHU 2375, and
Nicole Vogt for her rotation curve of GSS 294\_3364.  Thoughtful comments by a helpful, anonymous referee greatly improved our original
manuscript. We acknowledge the great cultural significance of Mauna Kea to native Hawaiians,
and express gratitude for permission to observe from its summit. Data presented herein were obtained at the W. M. Keck Observatory,
which is operated as a scientific partnership among the California Institute of Technology, the University of California, and
the National Aeronautics and Space Administration. The Observatory was made possible by the generous financial support of the W. M. Keck Foundation. This work was supported by the National Science
Foundation Science and Technology Center for Adaptive Optics, managed by the University of California at Santa Cruz under cooperative agreement No. AST-9876783, and by
the NSF grant to the DEEP survey AST-0071198. AJM appreciates support from the National Science Foundation from grant AST-0302153 through the NSF Astronomy and Astrophysics Fellows program. SAN was supported through STScI grant AR-08381.01-A.

\clearpage

\begin{deluxetable}{lcccccccccrcc}
\tablecaption{Journal of Keck AO Observations\label{table_journal}}
\tablewidth{0pt}
\tabletypesize{\tiny}
\rotate
\tablehead{& &\multicolumn{2}{c}{Coordinates (J2000.0)} & &\multicolumn{2}{c}{Guide Star} & &\multicolumn{2}{c}{PSF Star} & &\multicolumn{2}{c}{Exp.
Time
(s)}\\
\cline{3-4} \cline{6-7} \cline{9-10} \cline{12-13}
\colhead{Target} &\colhead{$z$} &\colhead{Right Asc.} &\colhead{Dec.} &
&\colhead{Mag. ($V$)} &\colhead{Offset ({{\arcsec}})} & &\colhead{Guide Mag. ($V$)} &\colhead{Offset ({{\arcsec}})} &\colhead{Date}
&\colhead{$H$} &\colhead{$K'$}}
\startdata
\objectname{JHU 2375} &0.531 &$00^{\rm h}17^{\rm m}13.88^{\rm s}$ &$15^{\circ}$46\arcmin15\farcs6 & &12.0 &25 & &12.0 &23 &21 August 2000
&\nodata &2400\\
\objectname{GSS 294\_3364} &0.651 &$14^{\rm h}15^{\rm m}24.62^{\rm s}$ &$52^{\circ}$01\arcmin54\farcs9 & &11.5 &29 & &10.0 &30 &7-9 May
2001 &2700 &\nodata\\
\objectname{GSS 294\_3367} &0.928 &$14^{\rm h}15^{\rm m}24.58^{\rm s}$ &$52^{\circ}$01\arcmin51\farcs5 & & &27 & & & & &3000 &2400\\
\enddata
\end{deluxetable}

\clearpage

\begin{deluxetable}{lccccccc}
\tablecaption{PSF FWHM in Arcseconds\label{table_resolution}}
\tablewidth{0pt}
\tablehead{& & &\multicolumn{2}{c}{$H$} & &\multicolumn{2}{c}{$K'$}\\
\cline{4-5} \cline{7-8}
\colhead{Target} &\colhead{$V$} &\colhead{$I$} &\colhead{Best} &\colhead{Worst} & &\colhead{Best} &\colhead{Worst}}
\startdata
\objectname{JHU 2375} &0.17 &0.20 &\nodata &\nodata & &0.13 &0.31\\
\objectname{GSS 294\_3364} &0.17 &0.20 &0.15 &0.24 & &\nodata &\nodata\\
\objectname{GSS 294\_3367} &0.17 &0.20 &0.15 &0.23 & &0.12 &0.23\\
\enddata
\end{deluxetable}

\clearpage

\begin{deluxetable}{lccccccccccccc}
\tablecaption{Model Structural Parameters\tablenotemark{1}\label{table_ratios}}
\tablewidth{0pt}
\tabletypesize{\tiny}
\rotate
\tablehead{& & & & & & & & &\multicolumn{2}{c}{$(B/T)_H$} & &\multicolumn{2}{c}{$(B/T)_{K'}$}\\
\cline{10-11} \cline{13-14}
\colhead{Target} &\colhead{$R_{\rm disk}$ (kpc)} &\colhead{$i$ (deg)} &\colhead{${\rm PA}_{\rm disk}$} &\colhead{$R_{\rm bulge}$ (kpc)}
&\colhead{$e_{\rm bulge}$} &\colhead{${\rm PA}_{\rm bulge}$}
&\colhead{$(B/T)_V$} 
&\colhead{$(B/T)_I$} &\colhead{Best} &\colhead{Worst} & &\colhead{Best} &\colhead{Worst}}
\startdata
\objectname{JHU 2375}      &$4.32^{+0.06}_{-0.06}$ &$60.5^{+0.4}_{-0.5}$ &$134^{+1}_{-1}$ &$<5.5$ &\nodata &\nodata &$<0.01$               
&$<0.01$                &\nodata &\nodata & &$<0.34$
&$<0.54$\\
\objectname{GSS 294\_3364} &$6.18^{+0.08}_{-0.08}$ &$77.2^{+0.3}_{-0.3}$ &$100^{+1}_{-1}$ &$4.2^{+0.5}_{-0.2}$ &$0.70^{+0.01}_{-0.01}$
&~$94^{+1}_{-1}$ &$0.08^{+0.02}_{-0.03}$ &$0.32^{+0.04}_{-0.07}$
&$0.50^{+0.07}_{-0.08}$
&$0.58^{+0.11}_{-0.10}$ & &\nodata &\nodata\\
\objectname{GSS 294\_3367} &$1.72^{+0.03}_{-0.04}$ &$64.0^{+1.2}_{-1.1}$ &$162^{+1}_{-1}$ &$<4.8$ &\nodata &\nodata &$<0.01$               
&$<0.01$                &$<1.00$ &$<1.00$ & &$<1.00$
&$<1.00$\\
\enddata
\tablenotetext{1}{Uncertainties are 99\% confidence limits.}
\end{deluxetable}

\clearpage

\begin{deluxetable}{lccccccccccccc}
\tablecaption{Galaxy Photometry\tablenotemark{1}\label{table_photometry}}
\tablewidth{0pt}
\tabletypesize{\tiny}
\rotate
\tablehead{& & &\multicolumn{2}{c}{$H$} & &\multicolumn{2}{c}{$K'$} & &\multicolumn{2}{c}{$I-H$} & &\multicolumn{2}{c}{$I-K'$}\\
\cline{4-5} \cline{7-8} \cline{10-11} \cline{13-14}
\colhead{Object} &\colhead{$V$} &\colhead{$I$} &\colhead{Best} &\colhead{Worst} & &\colhead{Best} &\colhead{Worst} &\colhead{$V-I$} &\colhead{Best}
&\colhead{Worst} & &\colhead{Best} &\colhead{Worst}}
\startdata
\sidehead{JHU 2375}
Total &$20.80^{+0.01}_{-0.02}$ &$20.27^{+0.03}_{-0.03}$ &\nodata &\nodata & &$16.97^{+0.02}_{-0.01}$ &\nodata &$0.53^{+0.04}_{-0.05}$ &\nodata
&\nodata & &$3.30^{+0.10}_{-0.04}$ &\nodata\\
Disk  &$20.80^{+0.01}_{-0.02}$ &$20.27^{+0.03}_{-0.03}$ &\nodata &\nodata & &$17.12^{+0.31}_{-0.13}$ &\nodata &$0.53^{+0.04}_{-0.05}$ &\nodata
&\nodata & &$3.15^{+0.34}_{-0.16}$ &\nodata\\
Bulge &$>23.3$                 &$>22.4$                 &\nodata &\nodata & &$>20.2$ &\nodata &\nodata &\nodata &\nodata & &\nodata &\nodata\\
Knot  &$22.68^{+0.06}_{-0.12}$ &$22.32^{+0.20}_{-0.20}$ &\nodata &\nodata & &$>19.2$ &\nodata &$0.36^{+0.26}_{-0.32}$ &\nodata &\nodata & &$<3.3$
&\nodata\\
\sidehead{GSS 294\_3364}
Total &$21.78^{+0.03}_{-0.03}$ &$20.07^{+0.02}_{-0.02}$ &$17.17^{+0.02}_{-0.05}$ &$16.34^{+0.04}_{-0.06}$ & &\nodata &\nodata
&$1.71^{+0.05}_{-0.05}$ &$2.90^{+0.04}_{-0.07}$ &$3.73^{+0.06}_{-0.08}$ & &\nodata
&\nodata\\
Disk  &$21.94^{+0.04}_{-0.05}$ &$20.50^{+0.07}_{-0.10}$ &$17.92^{+0.16}_{-0.16}$ &$17.30^{+0.32}_{-0.24}$ & &\nodata &\nodata
&$1.44^{+0.11}_{-0.15}$ &$2.58^{+0.23}_{-0.26}$ &$3.20^{+0.39}_{-0.34}$ & &\nodata
&\nodata\\
Bulge &$23.97^{+0.27}_{-0.18}$ &$21.31^{+0.26}_{-0.12}$ &$17.93^{+0.19}_{-0.15}$ &$16.93^{+0.21}_{-0.19}$ & &\nodata &\nodata
&$2.66^{+0.53}_{-0.46}$ &$3.38^{+0.45}_{-0.27}$ &$4.38^{+0.47}_{-0.31}$ & &\nodata
&\nodata\\
\sidehead{GSS 294\_3367}
Total &$23.44^{+0.09}_{-0.06}$ &$22.00^{+0.03}_{-0.04}$ &$18.78^{+0.18}_{-0.15}$ &$17.89^{+0.09}_{-0.16}$ & &$18.25^{+0.17}_{-0.12}$
&$17.32^{+0.12}_{-0.05}$ &$1.44^{+0.12}_{-0.10}$ &$3.22^{+0.21}_{-0.19}$
&$4.11^{+0.12}_{-0.20}$ & &$3.75^{+0.20}_{-0.16}$ &$4.68^{+0.15}_{-0.09}$\\
Disk  &$23.44^{+0.09}_{-0.06}$ &$22.00^{+0.03}_{-0.04}$ &$>20.0$ &$>19.8$        & &$>19.6$ &$>18.9$        &$1.44^{+0.12}_{-0.10}$ &$<2.0$ &$<2.4$   
&    &$<2.2$ &$<3.1$\\
Bulge &$>26.5$                 &$>23.8$                 &$18.78^{+0.18}_{-0.15}$ &$18.10^{+0.33}_{-0.32}$        & &$18.25^{+0.17}_{-0.12}$ &$17.63^{+0.37}_{-0.32}$        &\nodata                &\nodata &\nodata   
&    &\nodata &\nodata\\
\enddata
\tablenotetext{1}{Uncertainties are 99\% confidence limits. Bands are as observed at Earth}
\end{deluxetable}

\clearpage

\clearpage

\clearpage

\clearpage

\clearpage

\clearpage

\clearpage

\clearpage

\clearpage

\clearpage

\clearpage

\clearpage

\clearpage

\clearpage

\clearpage

\clearpage

\end{document}